\documentclass[aps,showpacs,amssymb,amsfonts,twocolumn,superscriptadress]{revtex4}
\usepackage{amsmath,bbm,mathrsfs}
\usepackage{graphicx}
\usepackage{amsfonts}
\usepackage{amssymb}

\def\textbf#1{{\bf #1}}
\def\be{\begin{equation}}
\def\ee{\end{equation}}
\def\ben{\begin{eqnarray}}
\def\een{\end{eqnarray}}
\def\eea{\end{array}}
\def\bea{\begin{array}}
\newcommand{\ot}[0]{\otimes}
\newcommand{\Tr}[1]{\mathrm{Tr}#1}
\newcommand{\bei}{\begin{itemize}}
\newcommand{\eei}{\end{itemize}}
\newcommand{\ket}[1]{|#1\rangle}
\newcommand{\bra}[1]{\langle#1|}

\newcommand{\proj}[1]{\ket{#1}\!\bra{#1}}

\begin{document}

\title{Towards measurable bounds on entanglement measures}

\author{Remigiusz Augusiak}\email{remigiusz.augusiak@icfo.es}\author{Maciej Lewenstein}
\affiliation{ICFO--Institute Ci\'encies Fot\'oniques,
Mediterranean Technology Park, 08860 Castelldefels (Barcelona),
Spain}

\begin{abstract}
While the experimental detection of entanglement provides already
quite a difficult task, experimental quantification of
entanglement is even more challenging, and has not yet been
studied thoroughly. In this paper we discuss several issues
concerning bounds on concurrence measurable collectively on copies
of a given quantum state. Firstly, we concentrate on the recent
bound on concurrence by Mintert--Buchleitner [F. Mintert and A.
Buchleitner, Phys. Rev. Lett. {\bf 98}, 140505 (2007)]. Relating
it to the reduction criterion for separability we provide yet
another proof of the bound and point out some possibilities
following from the proof which could lead to improvement of the
bound. Then, relating concurrence to the generalized robustness of
entanglement, we provide a method allowing for construction of
lower bounds on concurrence from any positive map (not only the
reduction one). All these quantities can be measured as mean
values of some two--copy observables. In this sense the method
generalizes the Mintert--Buchleitner bound and recovers it when
the reduction map is used. As a particular case we investigate the
bound obtained from the transposition map. Interestingly,
comparison with MB bound performed on the class of $4\ot 4$
rotationally invariant states shows that the new bound is positive
in regions in which the MB bound gives zero. Finally, we provide
measurable upper bounds on the whole class of concurrences.
\end{abstract}
\maketitle

\section{Introduction}
\label{intro}

Entanglement \cite{review-horo} is the property of quantum states
of multipartite systems that is absolutely crucial for the future
emerging quantum technologies: from quantum communication, through
quantum information, to quantum metrology and quantum sensing
\cite{roadmap}. For this reason there has been a considerable
interest in the recent years in designing feasible and efficient
methods of entanglement detection (see the recent review on this
subject \cite{GuhneToth}). Particularly important in this respect
are entanglement detection schemes which require only local
measurements of the entangled parts of the composite system. Among
the most important approaches to local entanglement detection the
following methods are perhaps the most popular and useful:
\begin{itemize}

\item {\bf Quantum state tomography}. This method employs
very many local measurements, and becomes impractical in higher
dimensions. It is very useful for low dimensional systems, where
frequently entanglement criteria for the states in question are
known \cite{white98,blatt1,blatt2}.

\item{\bf Entanglement "visibility" methods}. These require detection of
only some elements of the density matrix, but  for a continuous
family of measuring devices settings (cf. \cite{visibility}).

\item{\bf Bell's tests}. These methods do more than just a check of entanglement –-
they check also the non--locality of the quantum states
\cite{Bell}. Obviously, they do not detect the entangled states
that do not violate any Bell inequality \cite{Werner,lhv} (see
also the review \cite{reviewbell}).

\item {\bf Entanglement witnesses (EW)}.
These are observables that have positive averages on all separable
states, but a negative one on at least one entangled state
\cite{horo96,terhal}. They can be measured locally, and one can
optimize such measurements in various aspects \cite{guehnepra}.
Nowadays, entanglement witnesses are routinely used in experiments
to detect entanglement (see e.g. Refs.
\cite{blatt1,demartini,harald}).

\item {\bf "Direct" entanglement detection schemes}.
Such schemes have been proposed for pure \cite{huelga,Walborn} or
mixed states \cite{PHConc,MintertBuchleitner,RAMDPH}. Particularly
interesting are those using structural approximations of positive
maps \cite{PHorodecki_From,pawel_ekert,korbiczpra}.

\item{\bf "Nonlinear" entanglement witnesses}. Such objects involve measurement of several
copies typically. Examples are discussed, e.g., in Refs.
\cite{PHorodecki_From,guehne1}. There are related methods
employing measurements of variances \cite{guehne2} or higher order
correlation functions \cite{blatt2,korbicz}, or even entropic
uncertainty relations \cite{Takeuchi,guehne3,Julio_entr}.
\end{itemize}

With the exception of quantum tomography (which tries to get all
possible information about the state, but is very costly in
resources), all of the above listed methods aim at answering the
{\it qualitative} question: given a state, is it entangled? Only
in few instances, these methods allow for further characterization
of different classes of entanglement, and various kinds of
optimization. For example, EWs allow to distinguish different
classes of multipartite entanglement (cf. \cite{acin3}).
Optimization of EWs may concern their effectiveness (amount of
states detected), or the complexity of experimental implementation
\cite{hyllus}.

While, the qualitative entanglement detection problem is already
quite difficult and complex, it is even more challenging and
practically important to pose the quantitative question: {\it
given a state, how much entangled it is?}. For this aim one has to
use entanglement measures (EM). There are many of those (see e.g.
Refs. \cite{Michal-measures,Plenio-measures}), and there is no
canonical choice. Various measures are more, or less useful,
depending on the physical and quantum informational context.

Entanglement measures are typically non measurable directly.
Having chosen an EM, one needs reliable bounds on it, that can be
then directly measured experimentally (see Ref.
\cite{MintertAPB}). The new emerging area of {\it quantitative
entanglement detection} consists thus to a great extend in the
search for efficient and measurable bounds for EM's. This area is
the subject of this paper.

Recently, there were several other papers dealing with the problem
of efficient bounds on entanglement measures. Let us just recall
some of these achievements. One of the earliest results in this
area comes from Breuer \cite{BreuerBound}, who provided a lower
bound on concurrence employing a mean value of some particular
linear EW. Other interesting bound for concurrence of this sort
was proposed in the bipartite case by Mintert and Buchleitner
\cite{MintertBuchleitner} (see also Ref. \cite{Mintert} for the
bound in a more general fashion) and then generalized to the
multipartite scenario in Ref. \cite{Aolita}. Also, "dual" upper
bounds for concurrence in both the bipartite and multipartite
scenarios were provided in Ref. \cite{Guo}. Interestingly, all
these bounds are measurable collectively on two copies of a state
and the Mintert--Buchleitner (MB) bound was even very recently
measured \cite{Schmid} (see also Ref. \cite{Bovino} for another
experiment and \cite{Jaksch1} for proposal of an experiment in the
multipartite scenario both aiming at determination of the same
quantities, however, in purely qualitative context and also recent
criticism on these kinds of experiments \cite{Enk1,Enk2}). Let us
finally mention that the general approach to derivation of bounds
for various entanglement measures from an incomplete information
about the state (coming from the knowledge of averages of certain
observables) was recently worked out in a series of papers
\cite{Brandao,Brandao0,Plenio,Reimpell_1,Reimpell_2,Eisert,Wunderlich}.

The main purpose of the paper is presentation of a possible
generalization of the MB bound. Firstly, however, in Sec.
\ref{Preliminaries} we make a brief review in which we recall some
of the lower and upper bounds on concurrence mentioned above. In
particular we concentrate on the bounds measurable on two copies
of a state, i.e., those from Refs.
\cite{MintertBuchleitner,Mintert,Guo}, however, we recall also the
result of \cite{BreuerBound}. Then, in Sec. \ref{NewBounds} we
provide yet another proof of the Mintert--Buchleitner bound on
concurrence. Recently an alternative proof of the MB bound, based
on the upper bound on Uhlmann--Jozsa fidelity
\cite{UhlmannFid,Jozsa} from Ref. \cite{Miszczak}, was provided in
Ref. \cite{alternativeProof}. Here, utilizing the notion of the
conjugate function of concurrence, but also the bound on fidelity,
we provide yet another proof of MB bound and its multipartite
version from Ref. \cite{Aolita}. On the one hand, it seems that
this approach could give some possibilities of direct improvements
of the bound. On the other hand, the present approach allows to
relate the MB bound to the reduction map. This connection, in
turn, leads to a new method of derivation of lower measurable
bounds on concurrence from any positive map (Sec.
\ref{NewBounds}). All the obtained bounds are measurable on two
copies of a given state. Moreover, in the particular case of the
reduction map the method recovers the MB bound and therefore can
be treated as its generalization to the case of arbitrary positive
map. Particularly interesting is the transposition map, which is
known to be stronger in detection of entanglement than the
reduction one. Hence, as checked in the case of $4\ot 4$
rotationally invariant states, the resulting bound works for
states for which the MB bound does not. Unfortunately, surely due
to the lack of optimization, the bounding values are in general
much lower that the ones of the MB bound.

Further, in Sec. \ref{UpperBounds}, we extend a little the result
of Ref. \cite{Guo} showing that the whole class of concurrences
discussed in Refs. \cite{Sinolecka,Jap,Gour} may be bounded from
above by some measurable functions of the state. In Sec.
\ref{Conclusions} we conclude the paper pointing also out some
possibilities for further research.

\section{Recent lower and upper measurable bounds on concurrence}
\label{Preliminaries}
%

Before proceeding with detailed considerations let us recall the
definition of concurrence both in the bipartite and multipartite
scenario. In general, when dealing with the bipartite case we will
utilize the so--called $I$--concurrence introduced by Rungta {\it
et al.} \cite{Rungta} in an attempt to generalize the
Hill--Wootters concurrence \cite{Hill}. In the multipartite case
the extension of $I$--concurrence provided in Ref. \cite{Carvalho}
will be used. We will also recall some of the lower and upper
bounds on these concurrences measurable on two copies of a given
state, provided so far in the literature.

\subsection{Bipartite case}
\label{II.bipartite}

Let us start from the bipartite case. For this purpose let
$\ket{\psi_{AB}}$ denote some bipartite pure state from the
Hilbert space
$\mathcal{H}_{A}\ot\mathcal{H}_{B}=\mathbbm{C}^{d}\ot
\mathbbm{C}^{d}$. Following \cite{Rungta} we define
$I$--concurrence (hereafter called concurrence) of
$\ket{\psi_{AB}}$ as follows
\begin{eqnarray}\label{conc}
C(\ket{\psi_{AB}})&=&\sqrt{2\left(1-\Tr\varrho_{r}^{2}\right)}\\\nonumber
&=&\sqrt{4\sum_{i<j}\mu_{i}^{(r)}\mu_{j}^{(r)}},
\end{eqnarray}
when $\varrho_{r}$ stands for one of the reductions of
$\ket{\psi_{AB}}$, i.e., $\varrho_{r}=\Tr_{A(B)}\proj{\psi_{AB}}$
and $0\leq\mu_{i}^{(r)}\leq 1$ are eigenvalues of $\varrho_{r}$
(or squared Schmidt coefficients of $\ket{\psi_{AB}}$). Notice
that from the above definition it follows that for the maximally
entangled state from $\mathbbm{C}^{d}\ot \mathbbm{C}^{d}$, i.e.,
\begin{equation}\label{Maximally}
\ket{\psi_{+}^{(d)}}=\frac{1}{\sqrt{d}}\sum_{i=0}^{d-1}\ket{ii},
\end{equation}
the concurrence (\ref{conc}) is given by
$C(\ket{\psi_{+}^{(d)}})=\sqrt{2(d-1)/d}$. The extension of $C$ to
all mixed states acting on $\mathcal{H}_{A}\ot\mathcal{H}_{B}$ is
{\it via} the convex roof, which means that for any bipartite
mixed state $\varrho_{AB}$ one defines concurrence as
\begin{equation}
C(\varrho_{AB})=\min_{\{p_{i},\ket{\psi_{AB}^{(i)}}\}}\sum_{i}p_{i}C(\ket{\psi_{AB}^{(i)}}),
\end{equation}
where the minimum is taken over all such ensembles
$\{p_{i},\ket{\psi_{AB}^{(i)}}\}$ that
%
$\varrho_{AB}=\sum_{i}p_{i}\proj{\psi_{AB}^{(i)}}$.
%

Having recalled the definition of concurrence we may pass to the
lower bound provided in Ref. \cite{MintertBuchleitner} (from now
on we shall be omitting the subscripts $AB$). It was shown there
that the concurrence obeys the following inequality
\begin{equation}\label{boundMB}
C^{2}(\varrho)\geq 2 \max_{r=A,B}\left\{
\Tr\varrho^{2}-\Tr\varrho_{r}^{2}\right\}.
\end{equation}
In the case when the right--hand side is negative we put zero.

This result was further extended in Ref. \cite{Mintert}, where a
more general inequality for concurrence was shown, namely, for any
pair of bipartite density matrices $\varrho$ and $\sigma$ it holds
that
\begin{equation}\label{boundM}
C(\varrho)C(\sigma)\geq
2\max_{r=A,B}\{\Tr\varrho\sigma-\Tr\varrho_{r}\sigma_{r}\}.
\end{equation}
For $\sigma=\varrho$ one recovers inequality (\ref{boundMB}). What
is important and interesting about the bound (\ref{boundMB}) or
more generally about the bound (\ref{boundM}) is that both can be
determined as a mean value of some joint observable on the state
$\varrho\ot\sigma$ (two copies of $\varrho$ in the case of
(\ref{boundMB}). More precisely, the bound (\ref{boundM}) can be
rewritten as \cite{MintertBuchleitner,Mintert}:
\begin{equation}
C(\varrho)C(\sigma)\geq
\max_{r=1,2}\left\{\Tr\left(W_{r}\varrho\ot\sigma\right)\right\},
\end{equation}
where $W_{1}=4(P_{-}^{AA'}-P_{+}^{AA'})\ot P_{-}^{BB'}$ and
$W_{2}=4 P_{-}^{AA'}\ot (P_{-}^{BB'}-P_{+}^{BB'})$. Here and in
what follows by $P_{+}$ and $P_{-}$ we denote projectors onto
symmetric and antisymmetric subspace of the product
finite--dimensional Hilbert space $\mathcal{H}\ot\mathcal{H}$,
respectively, which are given by $(1/2)(\mathbbm{1}_{d^{2}}\pm
V^{(2)})$. Also, $\mathbbm{1}_{d}$ denotes a $d\times d$ unity
matrix and $V^{(2)}$ stands for the so--called swap operator, that
is, operator acting as
$V^{(2)}\ket{\varphi_{1}}\ket{\varphi_{2}}=\ket{\varphi_{2}}\ket{\varphi_{1}}$
for any pair of $\ket{\varphi_{1}}$ and $\ket{\varphi_{2}}$ from
$\mathcal{H}$. Superscripts $AA' (BB')$ are to indicate on which
of subsystems of $\varrho$ and $\sigma$ ($A$ and $B$ are
subsystems of $\varrho$ and $A'$ and $B'$ of $\sigma$) the
projectors $P_{\pm}$ act. Interestingly, experiments in which mean
values of these observables were achieved have been recently
performed \cite{Bovino,Schmid}. Note also that the efficiency of
the bound (\ref{boundMB}) was intensively investigated in Ref.
\cite{Plastino}.

Further, upper bound on $C$ "dual" to the bound (\ref{boundMB})
and also measurable on two copies of a given state was provided in
Ref. \cite{Guo}, where it was shown that
\begin{eqnarray}\label{boundDual}
C^{2}(\varrho)&\leq& 2\min_{r=A,B}
\left\{1-\Tr\varrho_{r}^{2}\right\}\nonumber\\&\equiv&
\min_{r=1,2}\left\{\Tr\left(\widetilde{W}_{r}\varrho^{\ot
2}\right)\right\}
\end{eqnarray}
with $\widetilde{W}_{1}=4P_{-}^{AA'}\ot \mathbbm{1}_{d^{2}}^{BB'}$
and $\widetilde{W}_{2}=4\mathbbm{1}_{d^{2}}^{AA'}\ot P_{-}^{BB'}$.

Let us now shortly discuss connection of the lower bound on the
concurrence (\ref{boundMB}) to the so--called entropic
inequalities introduced firstly in Ref. \cite{RHPHMH} (see also
Ref. \cite{RHPH}) as a next example of the so--called scalar
separability criteria. The entropic inequalities were further
developed in a series of papers
\cite{MHPH,Terhal_entr,VollbrechtWolf}. In the most common way
they can be written for any separable bipartite state $\varrho$ in
the form
\begin{equation}\label{entropic}
S_{\alpha}(\varrho)\geq S_{\alpha}(\varrho_{r}) \qquad (\alpha\geq
0,\;r=A,B),
\end{equation}
where by $S_{\alpha}$ we denoted the quantum Renyi entropy
$S_{\alpha}(\rho)=[1/(1-\alpha)]\log\Tr\rho^{\alpha}$. For
$\alpha\geq 1$ they simplify to the form
$\Tr\varrho_{r}^{\alpha}\geq \Tr\varrho^{\alpha}$ $(r=A,B)$. This,
after comparison with Eq. (\ref{boundMB}) means that the
right--hand side of (\ref{boundMB}) is positive if and only if the
entropic inequality (\ref{entropic}) with $\alpha=2$ is violated
for at least one of subsystems of $\varrho$. In other words, the
Mintert--Buchleitner bound works only for states of which
entanglement is detected by (\ref{entropic}) with $\alpha=2$.
However, as it follows e.g. from Refs. \cite{Abe,Tsallis,Batle}
(this was also indirectly confirmed in Ref. \cite{Plastino}), the
efficiency of the entropic inequalities with low $\alpha$s is
rather weak and grows significantly with $\alpha\to \infty$. Also,
what is particularly important any of the inequalities
(\ref{entropic}) cannot detect bound entanglement. The latter is a
consequence of the fact proven in Ref. \cite{VollbrechtWolf} that
the inequalities (\ref{entropic}) follow from the reduction
criterion in the sense that if for some $\varrho$ it holds that
$(I\ot R)(\varrho)\geq 0$, then the entropic inequalities
(\ref{entropic}) are satisfied for all $\alpha$. Here and in what
follows by $R$ we denote the so--called reduction map
\cite{MHPH,Cerf}, which is an example of positive but not
completely positive map of the form
\begin{equation}\label{reduction}
R(X)=\Tr(X)\mathbbm{1}_{d}-X \qquad (X\in M_{d}(\mathbbm{C})).
\end{equation}
The reduction map is decomposable\footnote{ We say that a given
map $\Lambda$ is decomposable if it can be written as
$\Lambda=\Lambda_{1}+\Lambda_{2}\circ T$, where $\Lambda_{1}$ and
$\Lambda_{2}$ are completely positive.} and therefore as such
cannot detect PPT entangled states. Consequently, the bound
(\ref{boundMB}) obviously cannot work for PPT entangled states.

On the other hand, as already mentioned, the good think about the
entropic inequalities is that as it was confirmed in a series of
works (see e.g. Refs. \cite{Abe,Tsallis,Batle}), they become
stronger in detection of entanglement for higher $\alpha$
Therefore it would be interesting to find relations between $C$ or
other entanglement measures and the inequalities
$\Tr\varrho_{r}^{\alpha}\geq \Tr\varrho^{\alpha}$ for higher
$\alpha$s than $2$. Such bounds would be stronger in the sense
that they would detect more entangled states and still for integer
$\alpha$ would be promising from the experimental point. The
latter is because they can be represented as a mean value of some
many--copy entanglement witness on $\alpha$ copies of a state
$\varrho$ \cite{PHorodecki_From}. Furthermore, it seems
interesting to connect $C$ to recent generalizations of entropic
inequalities for all positive maps \cite{RAJSPH,RAJS,RAJS2}. As
the generalized inequalities are able to detect bound
entanglement, one could have strong measurable bounds on
entanglement measures working also for PPT entangled states.

Let us finally briefly mention other approaches leading to
measurable lower bounds on concurrence and in general entanglement
measures. First, in Ref. \cite{BreuerBound} a general bound for
the concurrence as a straightforward continuation of the result of
Ref. \cite{ChenAlbeverioFei} (see also Ref. \cite{Datta} for some
extensions of these results). Specifically, it was shown that
\begin{equation}\label{boundBreuer}
C(\varrho)\geq \sqrt{\frac{2}{d(d-1)}}\,g(\varrho),
\end{equation}
where $g$ is any convex operator function\footnote{We say that $g$
is operator convex if it satisfies $g(p\rho_{1}+(1-p)\rho_{2})\leq
pg(\rho_{1})+(1-p)g(\rho_{2})$ for any pair of density matrices
$\rho_{1}$ and $\rho_{2}$ and probability $p$.} obeying
\begin{equation}\label{gfunction}
g(\proj{\psi})\leq 2\sum_{i<j}\sqrt{\mu_{i}\mu_{j}},
\end{equation}
for any pure state $\ket{\psi}$ with its Schmidt coefficients
$\sqrt{\mu_{i}}$. Examples of functions which satisfy both the
conditions and give measurable bounds on concurrence were provided
in Refs. \cite{BreuerBound,Julio}. In particular, Breuer in
\cite{BreuerBound} considered the function $g_{1}(\varrho)=-\Tr
(\varrho W)$ which is convex and showed that for
$\mathcal{W}_{\mathcal{V}}=d(I\ot
\Lambda_{\mathcal{V}})(P_{+}^{(d)})$ with $\Lambda_{\mathcal{V}}$
denoting the following positive map introduced in Ref.
\cite{BreuerMap} (see also Ref. \cite{Hall} for further
generalization of this map)
\begin{equation}\label{BreuerMap}
\Lambda_{\mathcal{V}}(X)=\Tr(X)\mathbbm{1}_{d}-X-\mathcal{V}X^{T}\mathcal{V}^{\dagger}
\end{equation}
with $\mathcal{V}$ standing for unitary antisymmetric (that is
$\mathcal{V}^{T}=-\mathcal{V}$) matrix with the only nonzero
elements $\pm 1$ lying on its anti--diagonal (see e.g. Ref.
\cite{BreuerMap}), the function $g_{1}$ also satisfies
(\ref{gfunction}). Specifically, this results in the bound
\begin{equation}\label{boundBreuerW}
C(\varrho)\geq
-\sqrt{\frac{2}{d(d-1)}}\,\Tr\left(\mathcal{W}_{\mathcal{V}}\varrho\right).
\end{equation}
Interestingly, in Sec. \ref{NewBounds} we generalize this
inequality to the case of any entanglement witness $W$ satisfying
$W\leq \mathbbm{1}_{d}$.

Let us conclude the section by noting that the general method
leading to lower bounds on entanglement measures which can be
obtained from mean values of some quantum observables was provided
recently in a series of papers
\cite{Brandao,Brandao0,Plenio,Reimpell_1,Reimpell_2,Eisert,Wunderlich}.

\subsection{Multipartite case}
%
Let us eventually pass to the multipartite scenario. Consider an
$N$--partite pure state $\ket{\psi}$ from some finite--dimensional
product Hilbert space
$\mathcal{H}^{(N)}=\mathcal{H}_{1}\ot\ldots\ot\mathcal{H}_{N}$.
Then, following Ref. \cite{Carvalho} we define concurrence of this
state as
\begin{equation}\label{boundMulti}
C^{(N)}(\ket{\psi})=2^{1-N/2}
\sqrt{2^{N}-2-\sum_{i}\Tr\varrho_{i}^{2}},
\end{equation}
where the sum runs over all subsystems of $\ket{\psi}$ (notice
that $\ket{\psi}$ has exactly $2^{N}-2$ proper subsystems). The
superscript $N$ is to emphasize that we deal with the multipartite
scenario ($C^{(2)}\equiv C$). Again, for mixed state the
concurrence $C^{(N)}$ is defined {\it via} the convex roof.

Utilizing the bipartite bound (\ref{boundMB}) it was shown in Ref.
\cite{Aolita} that for any pair of $N$--partite density matrices
$\varrho$ and $\sigma$, $C^{(N)}$ satisfies the following
inequality
\begin{eqnarray}\label{MultipartiteBoundLower}
C^{(N)}(\varrho)C^{(N)}(\sigma)&\geq&
\frac{4}{2^{N}}\left[\left(2^{N}-2\right)\Tr\varrho\sigma-
\sum_{S}\Tr\varrho_{S}\sigma_{S}\right]\nonumber\\
&=&\Tr\left(W^{(N)}\varrho\ot\sigma\right),
\end{eqnarray}
where the summation runs over all $2^{N}-2$ proper subsystems of
$\varrho$ and $\sigma$ (denoted here by $\varrho_{S}$ and
$\sigma_{S}$) and
\begin{equation}
W^{(N)}=4\left[\mathbf{P}_{+}-P_{+}^{(1)}\ot\ldots \ot
P_{+}^{(N)}-\left(1-2^{1-N}\right)\mathbf{P}_{-}\right].
\end{equation}
Here $\mathbf{P}_{+}$ ($\mathbf{P}_{-}$) denotes a projector onto
symmetric (antisymmetric) subspace of the Hilbert space
$\mathcal{H}^{(N)}\ot\mathcal{H}^{(N)}$, while analogously to the
bipartite case, $P_{+}^{(i)}$ $(P_{-}^{(i)})$ stands for a
projector onto symmetric (antisymmetric) subspace of
$\mathcal{H}_{i}\ot\mathcal{H}_{i}$, i.e., the Hilbert space
representing $i$th particles of $\varrho$ and $\sigma$.

In a similar manner to the bipartite case an upper bound for
$C^{(N)}$ dual to the above one (for $\sigma=\varrho$) was proved
in Ref. \cite{Guo}:
\begin{eqnarray}\label{MultipartiteBoundUpper}
\left[C^{(N)}(\varrho)\right]^{2}&\leq&
\frac{4}{2^{N}}\left[\left(2^{N}-2\right)-\sum_{S}\Tr\varrho_{S}^{2}\right]\nonumber\\
&=&\Tr\left(\widetilde{W}^{(N)}\varrho\ot\varrho\right)
\end{eqnarray}
with $\widetilde{W}^{(N)}=W^{(N)}+8(1-2^{1-N})\mathbf{P}_{-}$. It
follows straightforwardly from (\ref{MultipartiteBoundLower}) and
(\ref{MultipartiteBoundUpper}) that both bounds can be measured as
mean values of observables $W^{(N)}$ and $\widetilde{W}^{(N)}$,
respectively on two copies of $\varrho$ (or $\varrho\otimes\sigma$
in more general case in (\ref{MultipartiteBoundLower})).

%
\section{Proofs of the lower bounds on concurrence}
\label{Proofs}
%

At the very beginning we provide a new proof of the the inequality
(\ref{boundM}) and in particular the MB inequality
(\ref{boundMB}). For this aim we will utilize the notion of
conjugate function (see Refs. \cite{ConvexBook,Audenaert}) of
entanglement measures and in particular concurrence (this notion
was recently utilized in Refs. \cite{Reimpell_1,Reimpell_2,Eisert}
to derive measurable bounds on the entanglement measures from mean
values of quantum observables) and the very recent upper bound on
the fidelity \cite{UhlmannFid,Jozsa} proved by Miszczak {\it et
al.} \cite{Miszczak}. It has to be emphasized that an alternative
proof of the bounds (\ref{boundMB}) and (\ref{boundM}) basing on
the latter has been recently provided in Ref.
\cite{alternativeProof}. Here we present a little bit different
approach which, in our opinion, could lead, at least in some
particular cases, to improvements of the bounds.

Let $E$ denote some entanglement measure, then following Refs.
\cite{ConvexBook,Audenaert} we define conjugate function of $E$ as
\begin{equation}\label{Legendre}
\hat{E}(W)=\sup_{\varrho}\left[\Tr(\varrho W)-E(\varrho)\right],
\end{equation}
where $W$ can be any observable including entanglement witnesses.
Notice that in the case of convex $E$ it suffices to take the
above supremum only over pure states. Omitting supremum in Eq.
(\ref{Legendre}), we obtain the following inequality
\begin{equation}\label{ineq1}
E(\varrho)\geq \Tr (W\varrho)-\hat{E}(W)
\end{equation}
satisfied by any $\varrho$. Thus, having measured some observable
$W$ (which can also be an entanglement witness) on some state
$\varrho$, we can use its mean value to bound entanglement of
$\varrho$ from below (for more detailed analysis following the
above approach see Refs. \cite{Reimpell_1,Reimpell_2,Eisert}).

\subsection{The bipartite case}

In order to proceed with our proof of the bound let us introduce,
following Ref. [44] (see also Ref. [54]), the observable
\begin{equation}\label{witness}
W_{\sigma}^{R}=-\frac{2}{C(\sigma)}(I\ot R)(\sigma)
\end{equation}
depending on an arbitrary bipartite entangled state $\sigma$ (to
have the operators $W_{\sigma}^{R}$ well defined we need to assure
that $C(\sigma)>0$). By $R$ we denote the reduction map given by
Eq. (\ref{reduction}).

Let us now proceed with the proof. Substituting Eq.
(\ref{witness}) into Eq. (\ref{ineq1}) and putting $E=C$, we get
\begin{eqnarray}\label{estimation1}
C(\varrho)&\geq& -\frac{2}{C(\sigma)}\Tr[\varrho(I\ot
R)(\sigma)]-\hat{C}(W_{\sigma}^{R})\nonumber\\
&=&\frac{2}{C(\sigma)}\left[\Tr(\varrho\sigma)-\Tr(\varrho\sigma_{A}\ot
\mathbbm{1}_{d})\right]-\hat{C}(W_{\sigma}^{R})\nonumber\\
&=&\frac{2}{C(\sigma)}\left[\Tr(\varrho\sigma)-\Tr(\varrho_{A}\sigma_{A})\right]-
\hat{C}(W_{\sigma}^{R}),
\end{eqnarray}
where to get the first equality we used the definition of $R$,
while the second equality follows from the fact that $\Tr(\varrho
X\ot \mathbbm{1}_{d})=\Tr(\varrho_{A}X)$ for any $X$. The only
problem with proving the bound (\ref{boundMB}) is to show that
$\hat{C}(W_{\sigma}^{R})=0$ or, even better, that in general
$\hat{C}(W_{\sigma}^{R})\leq 0$. In our case we have
\begin{eqnarray}\label{derivation}
&&\hat{C}(W_{\sigma}^{R})=\sup_{\ket{\phi}}\left\{-\frac{2}{C(\sigma)}
\bra{\phi}(I\ot R)(\sigma)\ket{\phi}-C(\ket{\phi})\right\}\nonumber\\
&&=\frac{2}{C(\sigma)}\sup_{\ket{\phi}}\left\{\bra{\phi}\sigma\ket{\phi}-
\Tr(\phi_{A}\sigma_{A})-\frac{1}{2}C(\sigma)C(\ket{\phi})\right\},\nonumber\\
\end{eqnarray}
where $\phi_{A}=\Tr_{B}\proj{\phi}$ and $\sigma_{A}$ is a
reduction of $\sigma$ to the first subsystem. Now, we can
decompose $\sigma$ with an optimal ensemble
$\{p_{i},\ket{\psi_{i}}\}$ with respect to concurrence $C$.
Substituting this into Eq. (\ref{derivation}), we get
\begin{eqnarray}\label{estimation2}
&&\hat{C}(W_{\sigma}^{R})=\frac{2}{C(\sigma)}\nonumber\\
&&\times\sup_{\ket{\phi}}
\left\{\sum_{i}p_{i}\left[|\langle\phi|\psi_{i}\rangle|^{2}-\Tr\big(\phi_{A}\sigma_{A}^{(i)}\big)
-\frac{1}{2}C(\ket{\psi_{i}})C(\ket{\phi})\right]\right\}\nonumber\\
&&\leq \frac{2}{C(\sigma)}\nonumber\\
&&\times\sum_{i}p_{i} \sup_{\ket{\phi}}
\left\{|\langle\phi|\psi_{i}\rangle|^{2}-\Tr\big(\phi_{A}\sigma_{A}^{(i)}\big)-
\frac{1}{2}C(\ket{\psi_{i}})C(\ket{\phi})\right\}.\nonumber\\
\end{eqnarray}
%
%
We can utilize the aforementioned upper bound for the
Ulhmann--Jozsa fidelity \cite{UhlmannFid} defined as
$F(\rho_{1},\rho_{2})=\Tr\sqrt{\sqrt{\rho_{1}}\rho_{2}\sqrt{\rho_{1}}}$.
Namely, it was shown in Ref. \cite{Miszczak} that the inequality
\begin{equation}\label{boundFidelity}
F^{2}(\rho_{1},\rho_{2})\leq
\Tr(\rho_{1}\rho_{2})+\sqrt{1-\Tr\rho_{1}^{2}}\sqrt{1-\Tr\rho_{2}^{2}}
\end{equation}
holds for any pair of density matrices $\rho_{1}$ and $\rho_{2}$.
On the other hand we know from Ref. \cite{UhlmannFid} that
$F(\rho_{1},\rho_{2})=\max_{\ket{\varphi_{1}},\ket{\varphi_{2}}}|\langle\varphi_{1}|\varphi_{2}\rangle|$,
where the maximum is taken over all purifications $\ket{\psi_{1}}$
and $\ket{\psi_{2}}$ of $\rho_{1}$ and $\rho_{2}$, respectively.
This means that $|\langle\varphi_{1}|\varphi_{2}\rangle|\leq
F(\rho_{1},\rho_{2})$ for any pure states $\ket{\varphi_{1}}$ and
$\ket{\varphi_{2}}$ and their reductions, i.e.,
$\rho_{i}=\Tr_{A(B)}\proj{\varphi_{i}}$. Application of this
inequality to Eq. (\ref{boundFidelity}) leads us to a conclusion
that
\begin{equation}\label{FidIneq}
|\langle\varphi_{1}|\varphi_{2}\rangle|^{2}\leq
\Tr(\rho_{1}\rho_{2})+\frac{1}{2}C_{2}(\ket{\varphi_{1}})C_{2}(\ket{\varphi_{2}}).
\end{equation}
Notice that using different approach this inequality was also
proved in Ref. \cite{MintertBuchleitner}. Then, comparison of
(\ref{estimation2}) and (\ref{FidIneq}) allows us to infer that
$\hat{C}(W_{\sigma}^{R})\leq 0$, which in turn, after substitution
to Eq. (\ref{estimation1}) gives
\begin{equation}
C(\sigma)C(\varrho)\geq
2(\Tr\varrho\sigma-\Tr\varrho_{r}\sigma_{r})\qquad (r=A,B).
\end{equation}
This is exactly the inequality (\ref{boundM}) and in the
particular case when $\sigma=\varrho$ it gives (\ref{boundMB}).
The question which follows naturally from this analysis is if for
some classes of states $\hat{C}(W_{\sigma}^{R})<0$ and if in this
case $\hat{C}(W_{\sigma}^{R})$ would be measurable on copies of
$\sigma$. This, if true at least for some classes of states, would
obviously improve the bound (\ref{boundMB}). Below we provide two
classes of states for which the exact value of concurrence is
analytically determined and one can prove that the quantity
$\hat{C}(W_{\sigma}^{R})$ is zero in these cases.

First, we consider the two--qubit Bell diagonal states
\begin{equation}
\varrho_{\mathrm{BD}}=p_{1}\proj{\psi_{+}}+p_{2}\proj{\psi_{-}}+p_{3}\proj{\phi_{+}}+p_{4}\proj{\phi_{-}},
\end{equation}
where $p_{1}+p_{2}+p_{3}+p_{4}=1$ and $\ket{\psi_{\pm}}$, and
$\ket{\phi_{\pm}}$ denote the well--known Bell states given by
$\ket{\psi_{\pm}}=(1/\sqrt{2})(\ket{00}\pm\ket{11})$
and
$\ket{\phi_{\pm}}=(1/\sqrt{2})(\ket{01}\pm\ket{10}).$
Without any loss of generality we can also assume that $p_{1}\geq
p_{2}\geq p_{3}\geq p_{4}$. Then, one knows from Ref. \cite{Hill}
that concurrence of $\varrho_{\mathrm{BD}}$ is given by
$C(\varrho_{\mathrm{BD}})=\max\{0,2p_{1}-1\}$. Then, one may check
that in this case $\hat{C}(W_{\varrho_{\mathrm{BD}}}^{R})=0$ as
the state for which the supremum is achieved is $\ket{\psi_{+}}$.

Now, let us discuss the arbitrarily dimensional isotropic states
\begin{equation}
\varrho_{\mathrm{izo}}(f)=\frac{1-f}{d^{2}-1}\left(\mathbbm{1}_{d}-P_{+}^{(d)}\right)+f
P_{+}^{(d)}
\end{equation}
with $P_{+}^{(d)}$ denoting a projector onto the maximally
entangled state $\ket{\psi_{+}^{(d)}}$ defined by Eq.
(\ref{Maximally}) and
$f=\langle\psi_{+}^{(d)}|\varrho_{\mathrm{izo}}(f)|\psi_{+}^{(d)}\rangle$.
It was shown in Ref. \cite{Rungta1} that concurrence of this class
of states is given by
\begin{equation}
C(\varrho_{\mathrm{izo}}(f))=\left\{\begin{array}{ll} 0, &\hspace{0.5cm} f\leq 1/d\\
(d/(d-1))(f-1/d), &\hspace{0.5cm} 1/d\leq f\leq 1.
\end{array}\right.
\end{equation}
Then one straightforwardly verifies that for the isotropic states
$\hat{C}(W_{\varrho_{\mathrm{izo}}}^{R})=0$ and the pure state
realizing the supremum in the definition of $\hat{C}$ is
$\ket{\psi_{+}^{(d)}}$.

Let us finally discuss a little bit more general inequality than
(\ref{boundM}), being a lower bound for the quantity introduced by
Uhlmann and called $\Phi$--concurrence \cite{Uhlmann}. It was
introduced in order to calculate the Holevo capacity of quantum
channels and then thoroughly analyzed in the series of papers in
the case of rank two quantum channels (see e.g. Ref.
\cite{Uhlmann3} and references therein). Notice also that in Ref.
\cite{Hildebrand} the quantity was further generalized.

To recall the definition of $\Phi$--concurrence let $\Phi:
M_{d}(\mathbb{C})\to M_{d}(\mathbb{C})$ be some quantum channel
\footnote{We say that a linear map $\Phi:M_{d}(C)\to M_{d}(C)$ is
positive if $\Phi(A)\geq 0$ for any positive $A\in M_{d}(C)$. We
say that $\Phi$ is completely positive if the map $I_{n}\ot\Phi$
with $I_{n}$ denoting an identity acting on $M_{n}(C)$ is positive
for any $n$. Finally if $\Phi$ is completely positive and
preserves the trace we call it a quantum channel.}. Then the
$\Phi$--concurrence is defined for pure states in the following
way
\begin{equation}
C(\Phi;\ket{\psi})=\sqrt{2\left(1-\Tr\left[\Phi(\proj{\psi})\right]^{2}\right)}
\end{equation}
and for mixed states in a standard way {\it via} the convex roof.
One immediately notices that for $\Phi$ being just a partial trace
over one of the subsystems of $\ket{\psi}$, the above reproduces
the concurrence $C$ given in (\ref{conc}).

Let us now pass to the aforementioned bound. To prove it we can
utilize a nice property of the fidelity $F$. It was shown in Ref.
\cite{Nielsen} that the fidelity does not decrease after
application of any quantum operation (represented by completely
positive trace--preserving map). More precisely, for any pair of
quantum states $\varrho$ and $\sigma$ and for any quantum channel
$\Phi$ the inequality
\begin{equation}\label{boundFidelity2}
F(\varrho,\sigma)\leq F(\Phi(\varrho),\Phi(\sigma))
\end{equation}
is satisfied. Combination of inequalities (\ref{boundFidelity})
and the above one leads us to
\begin{eqnarray}\label{boundFidelity3}
F(\varrho,\sigma)&\leq&
\Tr\left[\Phi(\varrho)\Phi(\sigma)\right]\nonumber\\
&&+\sqrt{1-\Tr\left[\Phi(\varrho)\right]^{2}}\sqrt{1-\Tr\left[\Phi(\sigma)\right]^{2}}.
\end{eqnarray}
Now, following the same reasoning as in Ref.
\cite{alternativeProof}, however, with inequality
(\ref{boundFidelity3}) instead of (\ref{boundFidelity}), we get
\begin{equation}
C(\Phi;\varrho)C(\Phi;\sigma)\geq
2(\Tr(\varrho\sigma)-\Tr\left[\Phi(\varrho)\Phi(\sigma)\right]).
\end{equation}
%

\subsection{Multipartite case}
%
Using analogous reasoning to the one from the bipartite case we
can prove the bound (\ref{MultipartiteBoundLower}). For this
purpose we introduce the linear map
$R^{(N)}:\mathcal{B}(\mathcal{H}^{(N)})\to
\mathcal{B}(\mathcal{H}^{(N)})$ given by
\begin{eqnarray}
R^{(N)}(\varrho)&=&\sum_{S\subset I}[I_{S'}\ot R_{S}](\varrho)\nonumber\\
&=&\sum_{S}\Tr_{S}(\varrho)-(2^{N}-2)\varrho,
\end{eqnarray}
where $I=\{1,\ldots,N\}$, the sum runs over all proper subsets $S$
of $I$ (in other words all nontrivial subsystems of $\varrho$) and
$\Tr_{S}$ denotes partial trace over the subsystem of $\varrho$
represented by subset $S$ ($S'$ denotes $I\setminus S$). In other
words we apply the reduction map to all possible nontrivial
subsystems of $\varrho$ and then take the superposition of the
resulting outputs. As an illustrative example let us consider an
application of $R^{(3)}$ to some three--partite state $\varrho$.
This can be written as
\begin{eqnarray}
R^{(3)}(\varrho_{ABC})&=&\varrho_{A}\ot\mathbbm{1}_{BC}+\mathbbm{1}_{A}\ot\varrho_{A}\ot\mathbbm{1}_{C}
+\mathbbm{1}_{AB}\ot\varrho_{C}\nonumber\\
&&+\mathbbm{1}_{A}\ot\varrho_{BC}+\mathbbm{1}_{B}\ot\varrho_{AC}+\varrho_{AB}\ot\mathbbm{1}_{C}\nonumber\\
&&-6\varrho_{ABC}.
\end{eqnarray}

Then, in a full agreement with the bipartite case let us consider
an arbitrary $N$--partite state $\sigma$ with $C^{(N)}(\sigma)>0$
and introduce the following observable
\begin{equation}
W_{\sigma}^{R^{(N)}}=-\frac{2^{2-N}}{C^{(N)}(\sigma)}R^{(N)}(\sigma).
\end{equation}
\begin{widetext}
Denoting by $\{p_{i},\ket{\psi_{i}}\}$ the optimal ensemble of
with respect to C(N), we can write
\begin{eqnarray}\label{concIneq}
\hat{C}^{(N)}(W_{\sigma}^{R_{N}})&\leq&
\frac{2^{2-N}}{C^{(N)}(\sigma)}\sum_{i}p_{i} \sup_{\ket{\phi}}
\left\{\sum_{j}\left[|\langle\phi|\psi_{i}\rangle|^{2}-\Tr\left(\phi^{(j)}\psi_{i}^{(j)}\right)\right]-
\frac{C^{(N)}(\ket{\psi_{i}})C^{(N)}(\ket{\phi})}{2^{2-N}}\right\}\nonumber\\
&\leq & \frac{2^{2-N}}{C^{(N)}(\sigma)}\sum_{i}p_{i}
\sup_{\ket{\phi}}\left\{
\sum_{j}\sqrt{1-\Tr(\psi_{i}^{(j)})^{2}}\sqrt{1-\Tr(\phi^{(j)})^{2}}-
\frac{C^{(N)}(\ket{\psi_{i}})C^{(N)}(\ket{\phi})}{2^{2-N}}\right\}.
\end{eqnarray}
\end{widetext}
To prove that $\hat{C}^{(N)}(W_{\sigma}^{R_{N}})\leq 0$ it
suffices to apply the Cauchy--Schwarz--Bunyakowsky inequality.
More precisely, application of the latter to the sum appearing in
(\ref{concIneq}) gives
\begin{eqnarray}
&&\sum_{j}\sqrt{1-\Tr(\psi_{i}^{(j)})^{2}}\sqrt{1-\Tr(\phi^{(j)})^{2}}\leq\nonumber\\
&&\sqrt{2^{N}-2-\sum_{j}\Tr(\psi_{i}^{(j)})^{2}}\sqrt{2^{N}-2-\sum_{j}\Tr(\phi^{(j)})^{2}},\nonumber\\
\end{eqnarray}
which in turn after substitution to (\ref{concIneq}) finishes the
proof. Again the natural question is if for some classes of states
the quantity $\hat{C}^{(N)}(W_{\sigma}^{R_{N}})$ is less than zero
which would improve the bound.

\section{Measurable lower bounds on concurrence from positive maps}
\label{NewBounds}
%

Here, following the idea of relating the MB bound to reduction
map, we provide a method allowing for derivation of other lower
bounds on concurrence $C$ from any positive map. All the bounds
are also measurable on two copies of a given $\varrho$. We will
achieve this aim by connecting $C$ to the generalized robustness
of entanglement $R_{g}$. Comparison on the class of $4\otimes 4$
rotationally invariant states confirms that the new method can
lead to bounds which are applicable to states for which the
Mintert--Buchleitner bound is not.

\subsection{Connecting the concurrence and the generalized robustness of entanglement}

At the very beginning let us start by relating the concurrence and
the generalized robustness of entanglement. Specifically, in what
follows we will show that $C$ can be bounded by some function of
the latter. Let us then start from the definition of generalized
robustness of entanglement and recall some of its properties. It
is an entanglement measure introduced in Ref. \cite{Steiner} as a
generalization of robustness of entanglement given in Ref.
\cite{VidalTarrach} and defined for a given $\varrho$ as the
smallest $s$ for which there exists such other (possibly
entangled) state $\sigma$ that the following state
\begin{equation}
\varrho'=\frac{1}{1+s}(\varrho+s\sigma)
\end{equation}
is separable. Notice that the restriction that only separable
states $\sigma$ can be used in the above reproduces the definition
of the robustness of entanglement from Ref. \cite{VidalTarrach}.
In the case of pure states both the functions, i.e., the
generalized robustness of entanglement and the robustness of
entanglement were shown \cite{Steiner,VidalTarrach} to be given by
the following simple expression
\begin{eqnarray}\label{rob_pure}
R_{g}(\ket{\psi})=\left(\sum_{i}\sqrt{\mu_{i}}\right)^{2}-1
=2\sum_{i< j}\sqrt{\mu_{i}\mu_{j}},
\end{eqnarray}
where $\sqrt{\mu_{i}}$ are the Schmidt coefficient of
$\ket{\psi}$. On the other hand, it was shown in Ref.
\cite{Brandao} that the generalized robustness of entanglement can
be defined in terms of the witnessed entanglement \cite{Brandao0}.
More precisely, it was shown that
\begin{equation}
R_{g}(\varrho)=\max\left\{0,-\min_{W\leq
\mathbbm{1}_{d}}\Tr(W\varrho)\right\}.
\end{equation}
The advantage of this formulation is that the measurement of a
mean value of some entanglement witness $W$ satisfying $W\leq
\mathbbm{1}_{d}$ on a given state $\varrho$ provides
simultaneously a lower bound (possibly negative) on its
entanglement. In other words, for any entanglement witness $W\leq
\mathbbm{1}_{d}$, one has
\begin{equation}
R_{g}(\varrho)\geq -\Tr (W\varrho).
\end{equation}

Let us now relate $C$ and $R_{g}$. For this purpose we utilize the
relation (\ref{boundBreuer}). On the one hand it was shown in Ref.
\cite{Steiner} that $R_{g}$ is an operator convex function. On the
other hand it follows from (\ref{rob_pure}) that the generalized
robustness of entanglement satisfies the condition
(\ref{gfunction}). As a consequence $R_{g}$ can be used in
(\ref{boundBreuer}) and therefore we have
\begin{equation}\label{boundConc}
C(\varrho)\geq \sqrt{\frac{2}{d(d-1)}}\,R_{g}(\varrho).
\end{equation}
Equality in the above is achieved for instance for
$\ket{\psi_{+}^{(d)}}$. What follows from this inequality is that
for any entanglement witness satisfying $W\leq \mathbbm{1}_{d}$
one has
\begin{eqnarray}\label{boundConc2}
C(\varrho)&\geq&
-\sqrt{\frac{2}{d(d-1)}}\,\Tr(W\varrho)\nonumber\\
&\equiv& -\sqrt{\frac{2}{d(d-1)}}\langle W\rangle_{\varrho}.
\end{eqnarray}

Let us notice that the above inequality generalizes to some extent
the result of Ref. \cite{BreuerBound} (cf. end of Sec.
\ref{II.bipartite}). It was shown there that the above inequality
holds for some particular witness $\mathcal{W}_{\mathcal{V}}$
which is the one following from the map $\Lambda_{U}$ (see Eq.
(\ref{BreuerMap})). Here we provided a general relation between
concurrence and mean value of any entanglement witness satisfying
$W\leq \mathbbm{1}_{d}$. However, due to this constraint the
inequality (\ref{boundConc2}) does not fully reproduce the result
of Breuer for the witness $\mathcal{W}_{\mathcal{V}}$ (cf. Sec.
\ref{II.bipartite}). This is because, irrespective on the
dimension, the largest eigenvalue of $\mathcal{W}_{\mathcal{V}}$
is two and therefore $\mathcal{W}_{\mathcal{V}}$ does not fulfil
the above condition. Of course, for the purposes of the inequality
(\ref{boundConc2}) it suffices to take
$\mathcal{W}_{\mathcal{V}}/2$. This results in the bound on
concurrence which works for exactly the same states as the one
from Ref. \cite{BreuerBound}, however, is not that tight.

\subsection{Bounds}

Now we can discuss how the inequality (\ref{boundConc2}) can be
utilized to provide lower bounds on concurrence measurable on two
copies of $\varrho$. For this purpose we need to find appropriate
entanglement witnesses $W$. On the one hand, following the idea
laying behind the proof of the MB bound to use the reduction map
(\ref{reduction}), an attempt to use other positive maps seems
natural. On the other hand, to get nonlinear bounds which are
measurable on two copies of given $\varrho$ we are interested in
these witnesses which are state--dependent (in a sense that to
construct $W_{\varrho}^{\Lambda}$ we use the state of which
entanglement is to be bounded). Taking these two remarks into
account we introduce the witnesses of the form
\begin{equation}\label{witnesses}
W_{\varrho}^{\Lambda}=\alpha^{\Lambda}_{\varrho}(I\ot\Lambda)(\varrho),
\end{equation}
where $\Lambda$ denotes arbitrary positive map and
$\alpha_{\varrho}^{\Lambda}$ is some constant which we will
specify later. It is clear from the definition that the mean value
of $W_{\varrho}^{\Lambda}$ is nonnegative on any separable state.
To see this explicitly let us notice that for any separable state
$\sigma_{\mathrm{sep}}=\sum_{i}q_{i}\sigma_{A}^{(i)}\ot
\sigma_{B}^{(i)}$ and positive map $\Lambda$ it holds that
\begin{eqnarray}
\langle W_{\varrho}^{\Lambda}\rangle_{\sigma_{\mathrm{sep}}}
=\sum_{i}q_{i}\Tr\left[\varrho\left(\sigma_{A}^{(i)}\ot\Lambda^{\dagger}(\sigma_{B}^{(i)})\right)\right]
\geq 0,
\end{eqnarray}
where $\Lambda^{\dagger}$ denotes the dual map of $\Lambda$, i.e.,
its conjugate map with respect to the Hilbert--Schmidt scalar
product, i.e., such map that $\Tr (X^{\dagger}\Lambda(Y))=\Tr
(\Lambda^{\dagger}(X)Y)$ for any $X$ and $Y$. The inequality in
the above follows from the fact that the dual map
$\Lambda^{\dagger}$ of some positive map $\Lambda$ is also
positive and that $\Tr(AB)\geq 0$ for positive matrices $A$ and
$B$. Let us notice that similarly to the case of the witness
$W^{R}_{\varrho}$ (cf. (\ref{witness})) there is no sense to use
separable $\varrho$ or completely positive maps in construction of
$W_{\varrho}^{\Lambda}$ as in such case its mean value is
nonnegative for all, even entangled states. On the other hand, as
we will se below, entangled states $\varrho$ and positive but not
completely positive maps $\Lambda$ may lead to useful entanglement
witnesses.

Application of $W_{\varrho}^{\Lambda}$ to the bound on $C$ in Eq.
(\ref{boundConc2}) leads us to
\begin{eqnarray}\label{boundConc3}
C(\varrho)&\geq& -\sqrt{\frac{2}{d(d-1)}}\,\langle
W_{\varrho}^{\Lambda}\rangle_{\varrho}\nonumber\\
&=&-\alpha_{\varrho}^{\Lambda}\sqrt{\frac{2}{d(d-1)}}\Tr[(I\ot\Lambda)(\varrho)\varrho],
\end{eqnarray}
provided, however, that $W_{\varrho}^{\Lambda}\leq
\mathbbm{1}_{d}$. The second issue which should be addressed here
it that we want somehow to optimize the bound in the sense that we
want the values $-\langle W_{\varrho}^{\Lambda}\rangle_{\varrho}$
to be as high as possible for entangled states. To deal with these
two issues we can utilize the freedom we still have in the
constant $\alpha_{\varrho}^{\Lambda}$. For this purpose let us
notice that any positive $\Lambda$ acting on a finite--dimensional
matrix algebra can be written as $\Lambda=\Lambda_{1}-\Lambda_{2}$
with $\Lambda_{i}$ $(i=1,2)$ being some completely positive maps.
One of possible ways to get this decomposition is to go through
the Choi--Jamio\l{}kowski isomorphism \cite{Choi,Jamiolkowski}.
More precisely, one has to determine the so--called Choi (or
dynamical) (see Ref. \cite{Bengtsson}) matrix $(I\ot
\Lambda)(P_{+}^{(d)})$ and then find the completely positive maps
corresponding to the positive and negative parts of this matrix
(i.e., subspaces spanned by eigenvectors of $(I\ot
\Lambda)(P_{+}^{(d)})$ corresponding to its positive and negative
eigenvalues). Alternative example of such decomposition is the one
in which the completely positive map $\Lambda_{1}$ can be taken to
be the one proportional to
$\Lambda_{\mathrm{Tr}}(\cdot)=\Tr(\cdot)\mathbbm{1}_{d}$.
Specifically, any positive map $\Lambda:M_{d}(\mathbbm{C})\to
M_{d}(\mathbbm{C})$ can be written as
\begin{equation}\label{CanDecomp}
\Lambda=\xi\Lambda_{\mathrm{Tr}}-\Lambda_{2},
\end{equation}
where $\xi=d\lambda_{\max}$ with $\lambda_{\max}$ denoting the
maximal eigenvalue of $(I\ot \Lambda)(P_{+}^{(d)})$.

According to the condition that $W_{\varrho}^{\Lambda}\leq
\mathbbm{1}_{d}$ we need to assure that for a given $\Lambda$ and
entangled $\rho$ it holds that
$\alpha^{\Lambda}_{\varrho}\bra{\phi}(I\ot
\Lambda)(\varrho)\ket{\phi}\leq 1$ for any pure state
$\ket{\phi}\in \mathbbm{C}^{d}\ot \mathbbm{C}^{d}$. For this one
may take
$(\alpha_{\varrho}^{\Lambda})^{-1}=\sup_{\ket{\psi}}\bra{\psi}(I\ot
\Lambda)(\varrho)\ket{\psi}$. Rough but easier to perform
estimation shows, however, that to fulfil this condition it
suffices to put \footnote{By $\|\cdot\|$ we denote the operator
norm, which in the case of positive matrix $A\geq 0$ is just
maximal eigenvalue of $A$. }
$\alpha_{\varrho}^{\Lambda}=1/\|(I\ot\Lambda_{1})(\varrho)\|$.
This is because in such case
$\alpha^{\Lambda}_{\varrho}\bra{\phi}(I\ot
\Lambda)(\varrho)\ket{\phi}\leq
\alpha^{\Lambda}_{\varrho}\bra{\phi}(I\ot
\Lambda_{1})(\varrho)\ket{\phi}\leq
\alpha^{\Lambda}_{\varrho}\|(I\ot\Lambda_{1})(\varrho)\|= 1$,
where we utilized the mentioned decomposition of $\Lambda$ and
complete positivity of $\Lambda_{i}$ $(i=1,2)$. The above analysis
leads us to the conclusion that we can consider the following
entanglement witnesses
\begin{equation}
\widetilde{W}_{\varrho}^{\Lambda}=\frac{(I\ot
\Lambda)(\varrho)}{\|(I\ot\Lambda_{1})(\varrho)\|}\leq
\mathbbm{1}_{d}.
\end{equation}
Taking into account the particular decomposition
(\ref{CanDecomp}), the above can be rewritten as
\begin{equation}
\widetilde{W}_{\rho}^{\Lambda}=\frac{1}{\xi\|\varrho_{A}\|}(I\ot
\Lambda)(\rho),
\end{equation}
which after application to (\ref{boundConc2}) allows us to write
\begin{equation}\label{boundConc4}
C(\varrho)\geq
-\frac{1}{\xi\|\varrho_{A}\|}\sqrt{\frac{2}{d(d-1)}}\Tr\left[(I\ot\Lambda)(\varrho)\varrho\right].
\end{equation}
In general, even though the value of
$\Tr\left[(I\ot\Lambda)(\varrho)\varrho\right]$ may be in
principle determined experimentally independently of $\varrho$
(see e.g. Ref. \cite{RAJSPH} or description below), we still have
to know $\|\varrho_{A}\|$ to determine the bound. Therefore, we
need to have some knowledge (which is maximal eigenvalue of one of
subsystems of $\varrho$) about the state for which we want to
estimate experimentally the lower bound on $C$. What we would like
to have, however, is kind of 'black box' which when given a state
$\varrho$ returns a lower bound on its concurrence {\it without
any} knowledge about $\varrho$. On the other hand, one knows that
for quite huge class of states the value of $\|\varrho_{A}\|$ is
constant. This is the class of states with at least one maximally
mixed subsystem $\varrho_{r}=\mathbbm{1}_{d}/d$ ($r=A$ or $r=B$).
For such class of states the inequality (\ref{boundConc4}) gives
\begin{equation}
C(\varrho)\geq
-\frac{1}{\xi}\sqrt{\frac{2d}{d-1}}\,\Tr\left[(I\ot\Lambda)(\varrho)\varrho\right].
\end{equation}

Let us now discuss this construction in the case of a particular
positive map, namely, the transposition
$T_{U}(X)=UX^{T}U^{\dagger}$ with $U$ being some unitary matrix.
According to the decomposition (\ref{CanDecomp}), transposition
can be expressed as
$T_{U}=\Lambda_{\mathrm{Tr}}-(\Lambda_{\mathrm{Tr}}-T_{U})$ (here
$\xi=1$), where $\Lambda_{\mathrm{Tr}}-T_{U}$ may be easily shown
to be completely positive as after normalization it is just the
Werner--Holevo channel \cite{WernerHolevo}. For states with at
least one maximally mixed subsystem, say
$\varrho_{A}=\mathbbm{1}_{d}/d$, we see that
$\widetilde{W}_{\varrho}^{T_{U}}=d \varrho^{\Gamma_{U}}$, where
$\Gamma_{U}=I\ot T_{U}$ denotes the partial transposition acting
on the subsystem $B$. Putting this to Eq. (\ref{boundConc2}) we
get that
\begin{equation}\label{boundTransp}
C(\varrho)\geq
-\sqrt{\frac{2d}{d-1}}\,\Tr\left(\varrho\varrho^{\Gamma_{U}}\right).
\end{equation}
Notice that we still have some freedom in the choice of the
unitary matrix $U$ and thus we can always optimize over it.
However, this makes the measurability of the bound
state--dependent.

In a similar way we can consider other positive maps than
transposition. For instance, we can analyze the reduction map,
which gives the inequality $C(\varrho)\geq
\sqrt{2d/(d-1)}(\Tr\varrho^{2}-\Tr\varrho_{r}^{2})$ $(r=A,B)$.
However, surely because of the lack of optimization, this
inequality is in general weaker than (\ref{boundMB}) in the sense
that its right--hand side gives lower values (more precisely, this
is because in the MB bound one has a square root of the term
$\Tr\varrho^{2}-\Tr\varrho_{r}^{2}$ which is always lower than
one). On the other hand, in the case of the reduction map we can
be a little bit more clever. This is because in this case we have
the inequality (\ref{boundFidelity}) and therefore we can propose
a bit "better" entanglement witness. Specifically, we can choose
\begin{equation}
\mathscr{W}^{R}_{\varrho}=-\sqrt{\frac{d}{2(d-1)}}\frac{1}{C(\varrho)}(I\ot
R)(\varrho),
\end{equation}
i.e., we can put
$\alpha_{\varrho}^{R}=\sqrt{d/2(d-1)}[1/C(\varrho)]$ in Eq
(\ref{witnesses}). Then, utilizing (\ref{boundFidelity}) we can
prove that $\mathscr{W}^{R}_{\varrho}\leq \mathbbm{1}_{d}$ meaning
that this witness can be utilized in (\ref{boundConc3}), which
gives finally the MB bound. It should be stresses that in this way
we get another proof of the inequality (\ref{boundMB}).

Let us finally shortly discuss the way in which the value of
$\Tr[(I\ot \Lambda)(\varrho)\varrho]$ and in particular
$\Tr(\varrho \varrho^{\Gamma_{U}})$ may be determined as mean
values of some two--copy observables. The general approach was
worked out in Ref. \cite{RAJSPH} as generalization of Ref.
\cite{PHorodecki_From} (see also Refs. \cite{PHConc,pawel_ekert}).
For this purpose we utilize the swap operator $V^{(2)}$ already
introduced in Sec. \ref{II.bipartite}, however this time we assume
that it permutes pure states belonging to some general
$N$--partite Hilbert space $\mathcal{H}^{(N)}$. One can easily
check that for any two Hermitian operators $A$ and $B$ acting on
$\mathcal{H}^{(N)}$ it holds that $\Tr(V^{(2)}A\ot B)=\Tr AB$
\cite{Isham}. Thus, introducing the notation
$\Theta_{\Lambda}(\varrho^{\ot
2})=(I\ot\Lambda)(\varrho)\ot\varrho$, we can write that
$\Tr[(I\ot\Lambda)(\varrho)\varrho]=\Tr\{V^{(2)}[(I\ot\Lambda)(\varrho)\ot\varrho]\}=
\Tr[V^{(2)}\Theta_{\Lambda}(\varrho^{\ot
2})]=\Tr[\Theta_{\Lambda}^{\dagger}(V^{(2)})\varrho^{\ot 2}]$,
where $\Theta_{\Lambda}^{\dagger}$ denotes the dual map of
$\Theta_{\Lambda}$. Since the map $\Lambda$ is positive it is also
Hermiticity preserving and thus as $V^{(2)}$ is Hermitian, the
resulting operator $\Theta_{\Lambda}^{\dagger}(V^{(2)})$ is also a
Hermitian one. Thus we can treat it as some collective two--copy
observable
$\mathcal{O}_{\Lambda}=\Theta_{\Lambda}^{\dagger}(V^{(2)})$ which
depends only on the positive map $\Lambda$ and not on the input
state. As a result we have that
$\Tr[(I\ot\Lambda)(\varrho)\varrho]$ can be expressed as a mean
value of some observable on two copies of a given $\varrho$, i.e.,
\begin{equation}
\Tr[(I\ot\Lambda)(\varrho)\varrho]=\Tr\left(\mathcal{O}_{\Lambda}\varrho^{\ot
2}\right).
\end{equation}

The particular case of the above was already discusses in Ref.
\cite{RAJSPH} for the partial transposition with respect to an
arbitrary subsystem of $N$--qubit $\varrho$ or even the full
transposition (this is useful in measurement of concurrence). Let
$I$ and $I'$ denote the set of all parties and the parties of
$\varrho$ on which we perform the transposition, respectively.
Then let $\Lambda=\tau^{I'}$ denote the partial transposition with
respect to the parties $I'$ followed by local rotation with the
second Pauli matrix $\sigma_{y}$. In this case the observable
$\mathcal{O}_{\tau^{I'}}$ is of the form \cite{RAJSPH}:
\begin{equation}
\mathcal{O}_{\tau^{I'}}=2^{|I'|}\bigotimes_{i\in I\setminus
I'}V^{(2)}_{A_{i}A_{i}'}\bigotimes_{i\in I'}P_{A_{i}A_{i}'}^{(-)},
\end{equation}
where $V^{(2)}_{A_{i}A_{i}'}$ stands for the swap operator
permutating $i$th parties of both copies of $\varrho$ (cf.
\ref{II.bipartite}) and $P_{A_{i}A_{i}'}^{(-)}$ is the projector
onto the antisymmetric subspace of the Hilbert space corresponding
to $i$th particles of both copies of $\varrho$.

\subsection{Comparison and effectiveness}

Let us now go back to inequality (\ref{boundTransp}) and compare
it to the Mintert--Buchleitner bound. Generally, surely because of
lack of optimization, the bound (\ref{boundTransp}) gives lower
values than (\ref{boundMB}). For instance, for the maximally
entangled state $P_{+}^{(d)}$ it is straightforward to show that
maximal value achievable by the right--hand side of
(\ref{boundTransp}) is $\sqrt{2/[d(d-1)]}$. To see this explicitly
it suffices to notice that the operator $P_{+}^{(d)\Gamma_{U}}$
has eigenvalues $\pm 1/d$. After comparison to (\ref{conc}) one
sees that this value is much smaller than exact value of
concurrence for $P_{+}^{(d)}$ which is
$C(P_{+}^{(d)})=\sqrt{2(d-1)/d}$ (recall that this value is
reproduced by the MB bound). However, as we shall see below, the
advantage of the bound (\ref{boundTransp}) is that there exist
states for which the MB bound gives zero, while at least for some
particular $U$ the bound (\ref{boundTransp}) a positive value and
simultaneously is nonlinear in $\varrho$. This is because, as
shown in Ref. \cite{RAJSPH} in the case of $4\otimes 4$
rotationally invariant bipartite states, the inequality
$\Tr(\varrho\varrho^{\Gamma_{U}})\geq 0$ (which can also be
treated as a criterion for separability) with $U=\mathcal{V}$,
where $\mathcal{V}$ being $4\times 4$ antisymmetric matrix with
only nonzero elements $\pm 1$ lying on its anti--diagonal, detects
entanglement of states which are not detected by the entropic
inequality (\ref{entropic}) with $\alpha=2$.

Let us now discuss the effectiveness of the presented
inequalities. For this purpose we apply the bound
(\ref{boundTransp}) to the aforementioned class of $4\ot 4$
$SO(3)$--invariant states and compare it to the bounds
(\ref{boundMB}). For the sake of completeness we also compare it
to the bound (\ref{boundBreuerW}). This class of states can be
represented as
\begin{equation}\label{SO3}
\varrho_{\mathrm{rot}}(p,q,r)=pP_{0}+qP_{1}+rP_{2}+(1-p-q-r)P_{3},
\end{equation}
where $p,q,r\geq 0$ and $p+q+r\leq 1$, and $P_{J}$ denote the
projector onto the eigenspace of the squared total angular
momentum divided by $2J+1$ (here the range of $J$ is
$J=0,\ldots,3$).

\begin{widetext}

\begin{figure}[h!]
(a)\includegraphics[width=4cm]{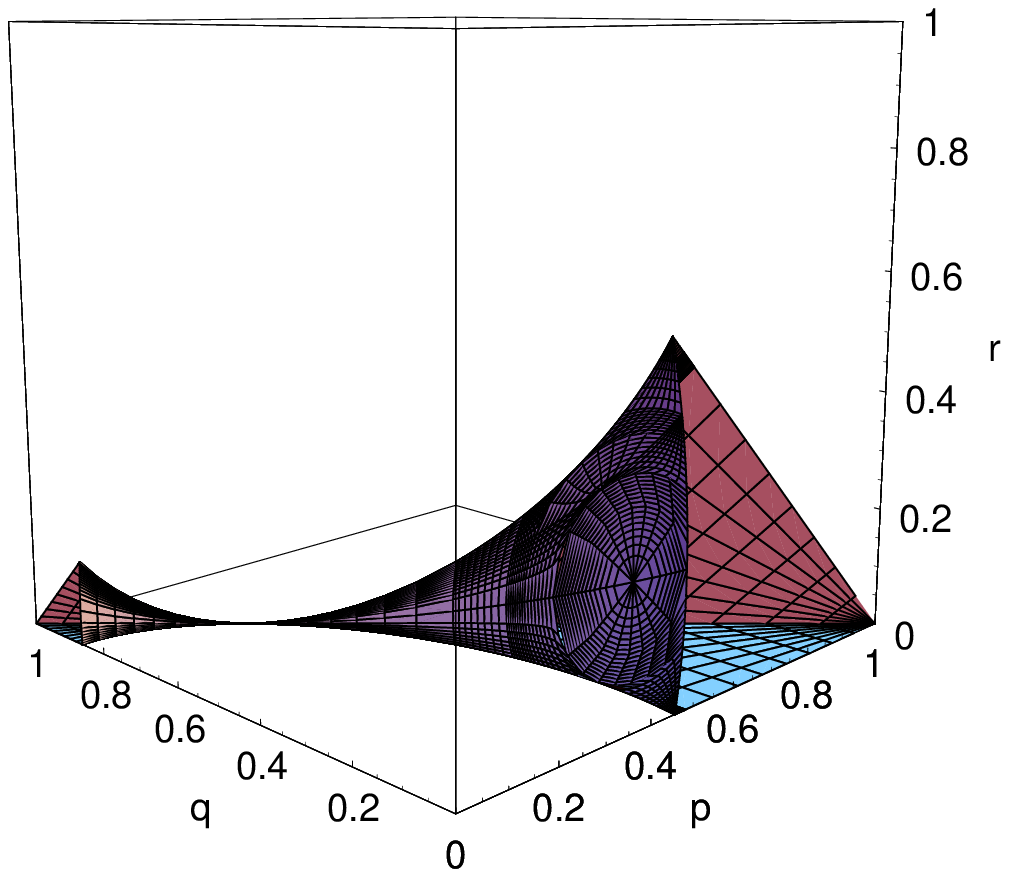}
(b)\includegraphics[width=4cm]{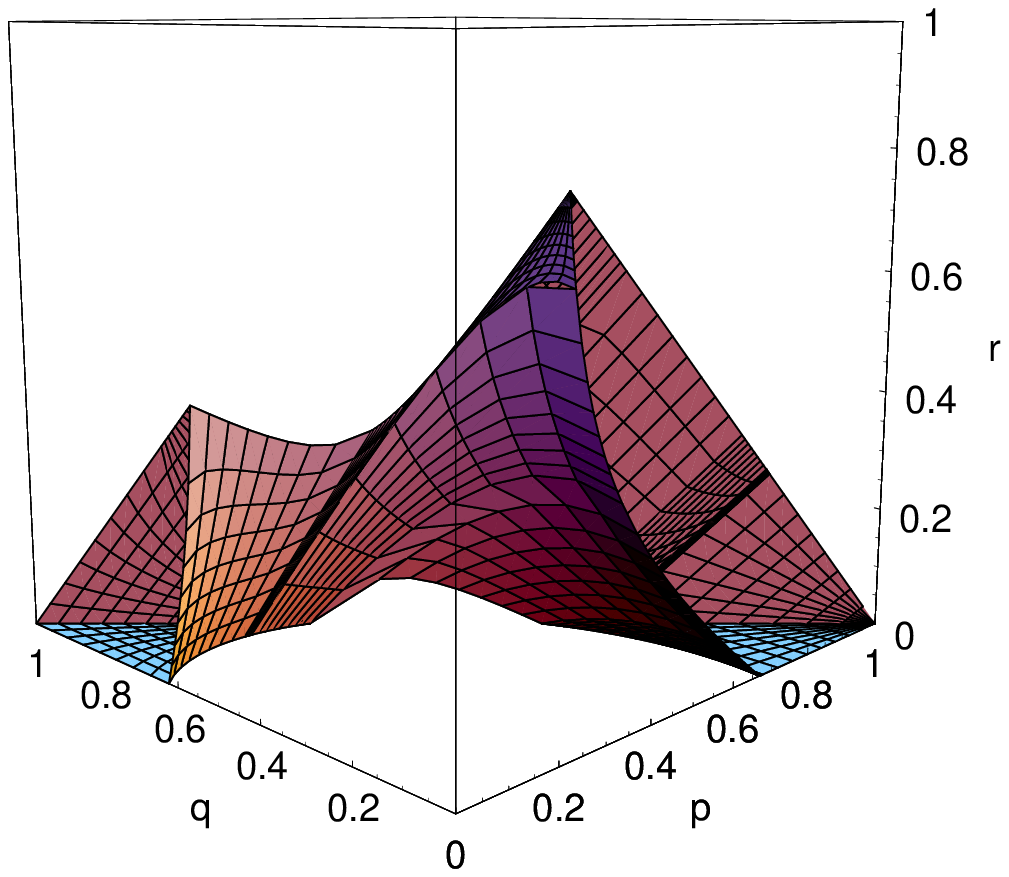}\\
(c)\includegraphics[width=4cm]{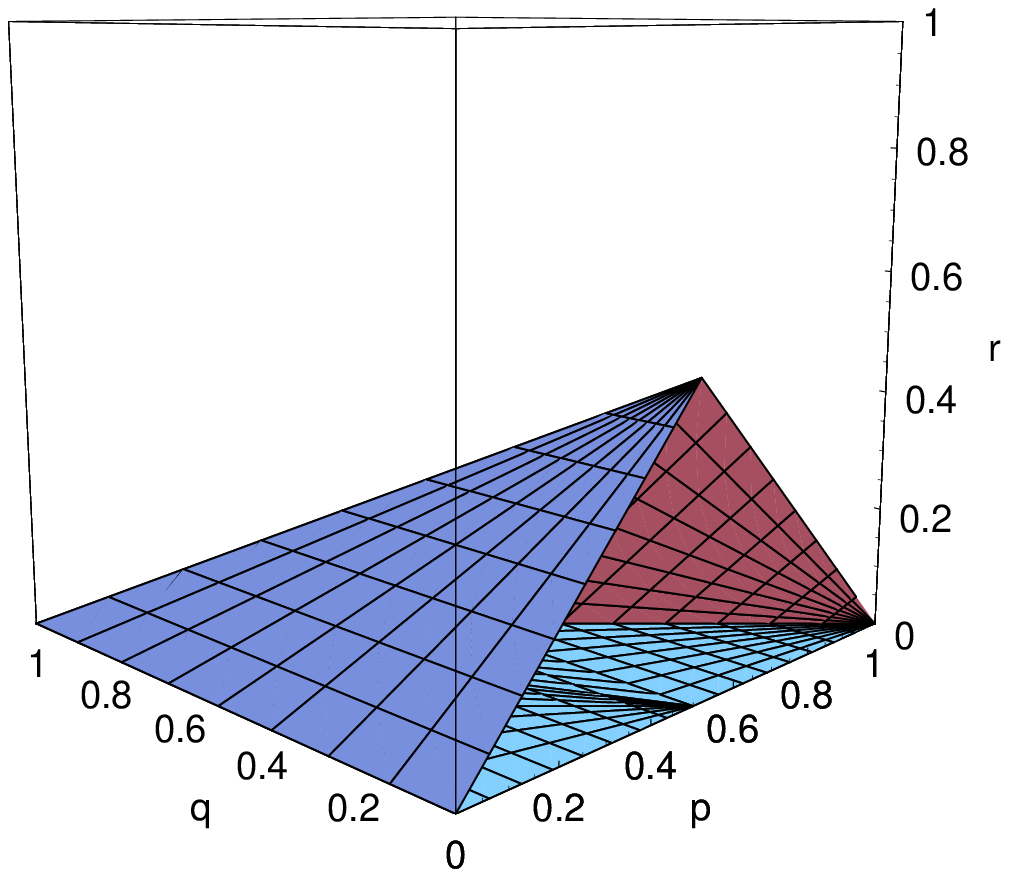}
(d)\includegraphics[width=4cm]{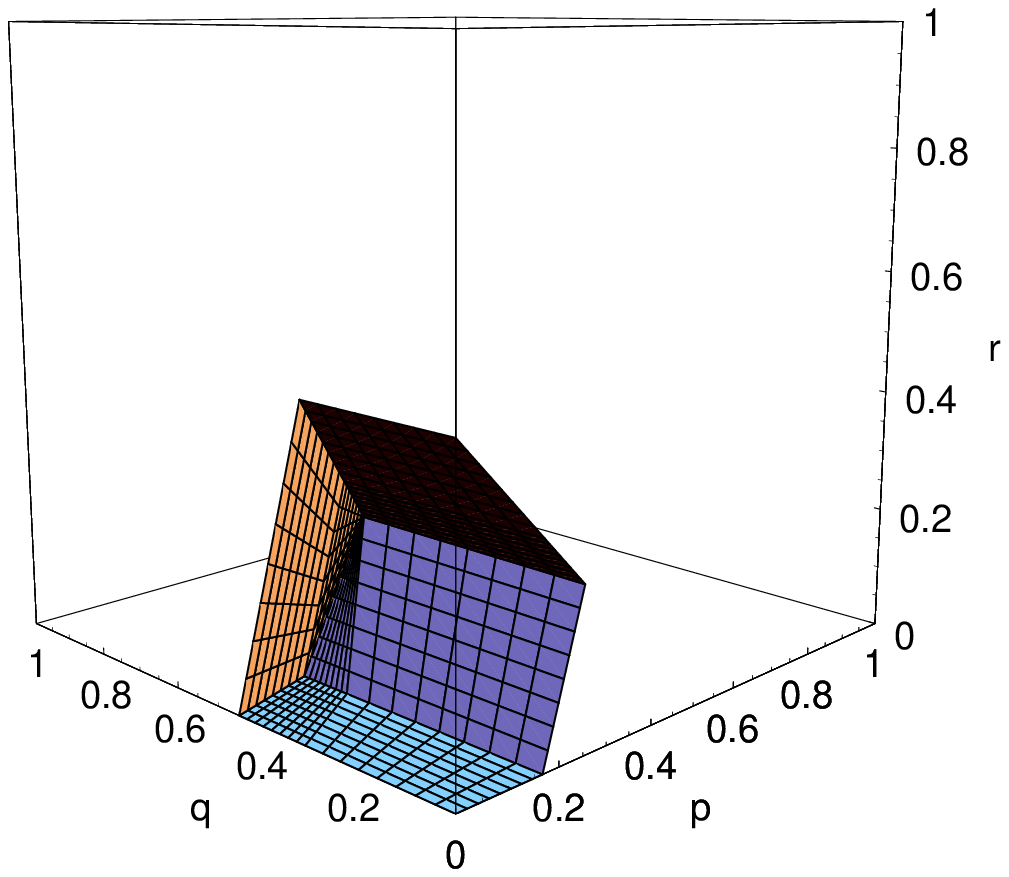}
\caption{Comparison of the regions of $4\ot 4$ rotationally
invariant states (\ref{SO3}) (remember that $p+q+r\leq 1$)
detected by three separability criteria lying behind the bounds
(\ref{boundMB}), (\ref{boundBreuerW}), and (\ref{boundTransp}).
Plot (a) is made for the entropic inequality (\ref{entropic}) with
$\alpha=2$, which corresponds to the bound (\ref{boundMB}) in the
sense that the latter works iff the former detects entanglement.
Plot (b) is made for the separability criterion
$\Tr(\varrho\varrho^{\Gamma_{U}})\geq 0$ (with $U$ being $4\times
4$ antisymmetric matrix with only nonzero elements $\pm 1$ lying
on its anti--diagonal) appearing in the bound (\ref{boundTransp}).
Finally, plot (c) shows region detected by the entanglement
witness $\mathcal{W}_{\mathcal{V}}$ appearing in the bound
(\ref{boundBreuerW}).
For completeness plot (d) presents the region of PPT $4\ot 4$
rotationally invariant states.} \label{FigDetection}
\end{figure}
\end{widetext}

In Fig. \ref{FigDetection} one finds comparison of three different
criteria for separability, lying behind the bounds
(\ref{boundMB}), (\ref{boundBreuerW}), and (\ref{boundTransp}).
Fig. \ref{FigDetection}(a) presents the region of parameters
$p,q,r$ in which the states $\varrho_{\mathrm{rot}}(p,q,r)$ are
detected by the entropic inequality (\ref{entropic}) with
$\alpha=2$. The latter is equivalent to the MB bound in the sense
that the MB bound works only for states which it detects. Thus
Fig. \ref{FigDetection}(a) presents subset of $4\otimes 4$
rotationally invariant states for which the bound (\ref{boundMB})
works. Similarly, Fig. \ref{FigDetection}(b) and Fig.
\ref{FigDetection}(c) present subsets of states which are detected
by the inequality $\Tr(\varrho\varrho^{\Gamma_{\mathcal{V}}})\geq
0$ and the entanglement witness $\mathcal{W}_{\mathcal{V}}$,
respectively. These are the states for which the bounds
(\ref{boundTransp}) and (\ref{boundBreuerW}) give a positive
value, respectively. Comparison of these two regions assures that
the new bound (\ref{boundTransp}) works for states for which the
MB bound (as well as the bound (\ref{boundBreuerW})) gives zero.

\begin{widetext}

\begin{figure}[h!]
(a)\includegraphics[width=5cm]{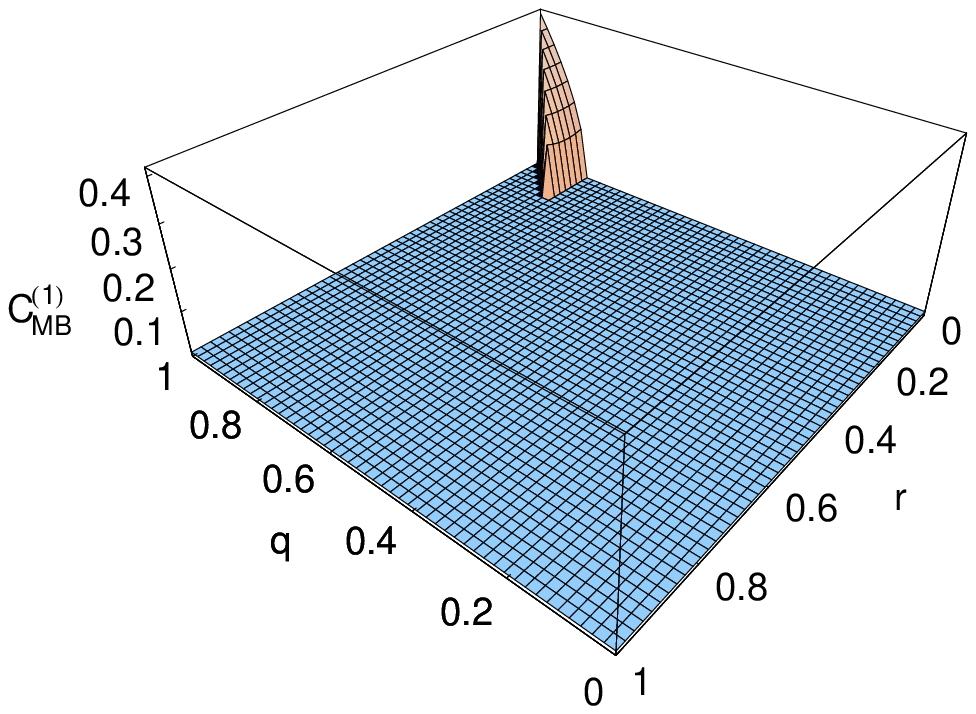}\hspace{0.5cm}\includegraphics[width=5cm]{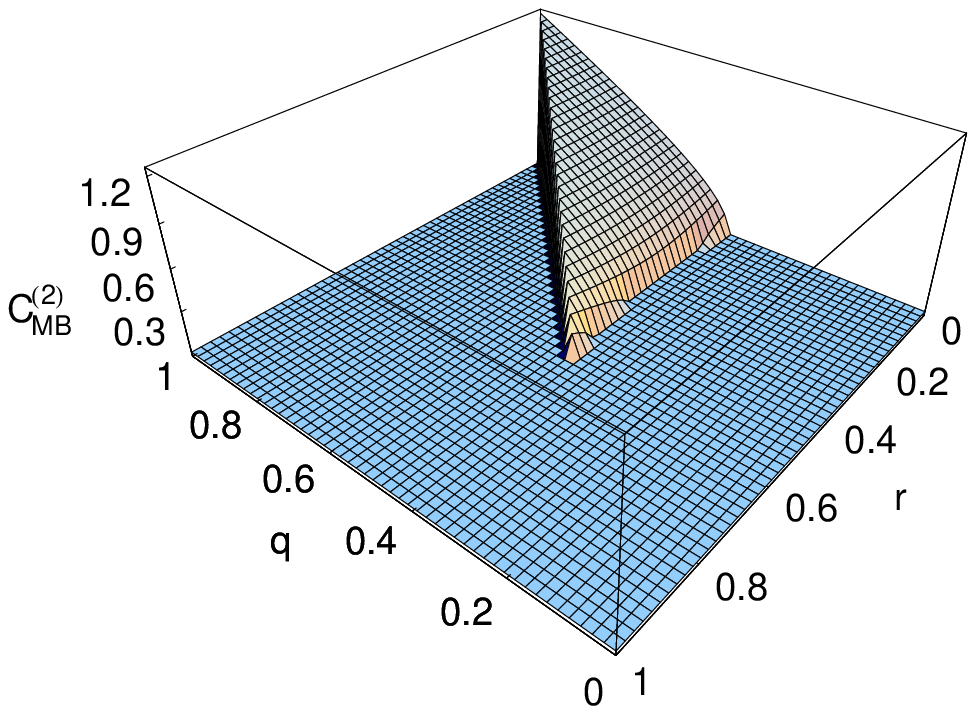}
\\
(b)\includegraphics[width=5cm]{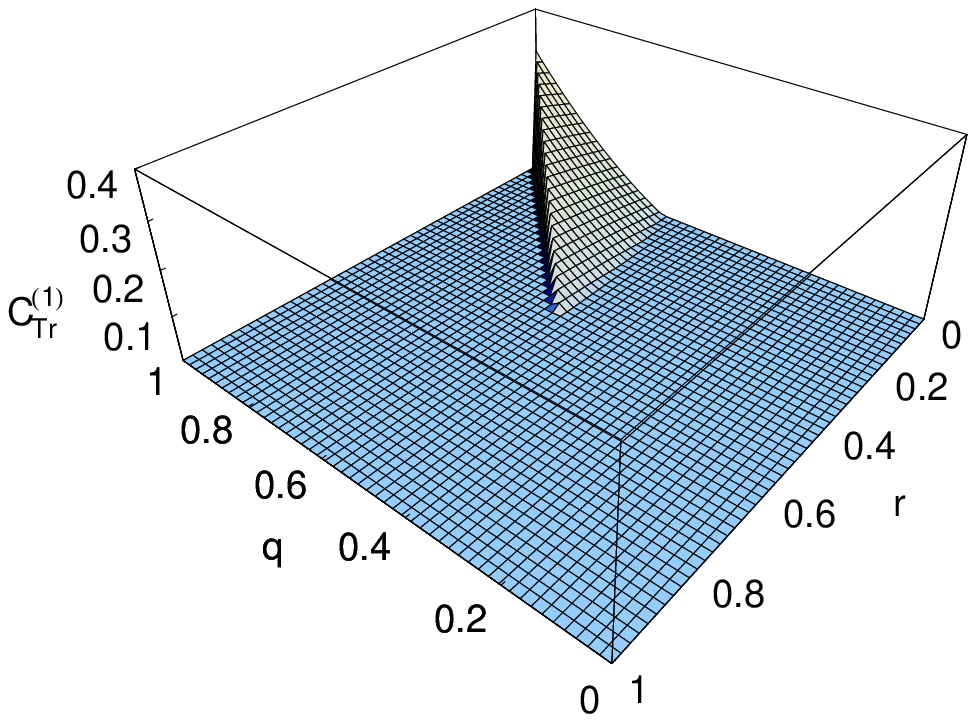}\hspace{0.5cm}
\includegraphics[width=5cm]{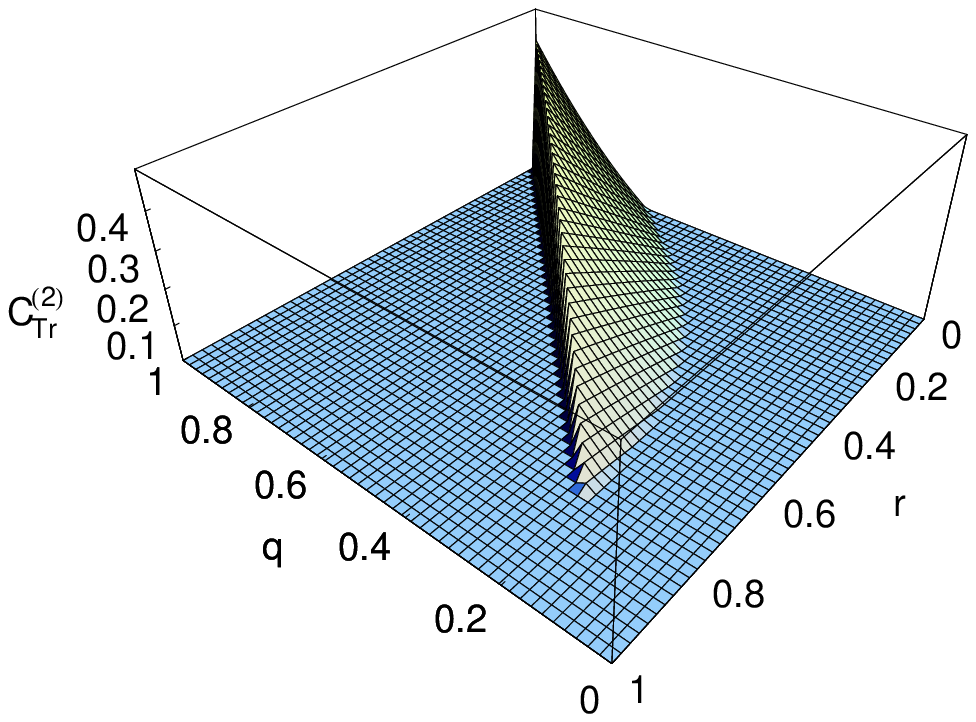}\\
(c)\includegraphics[width=5cm]{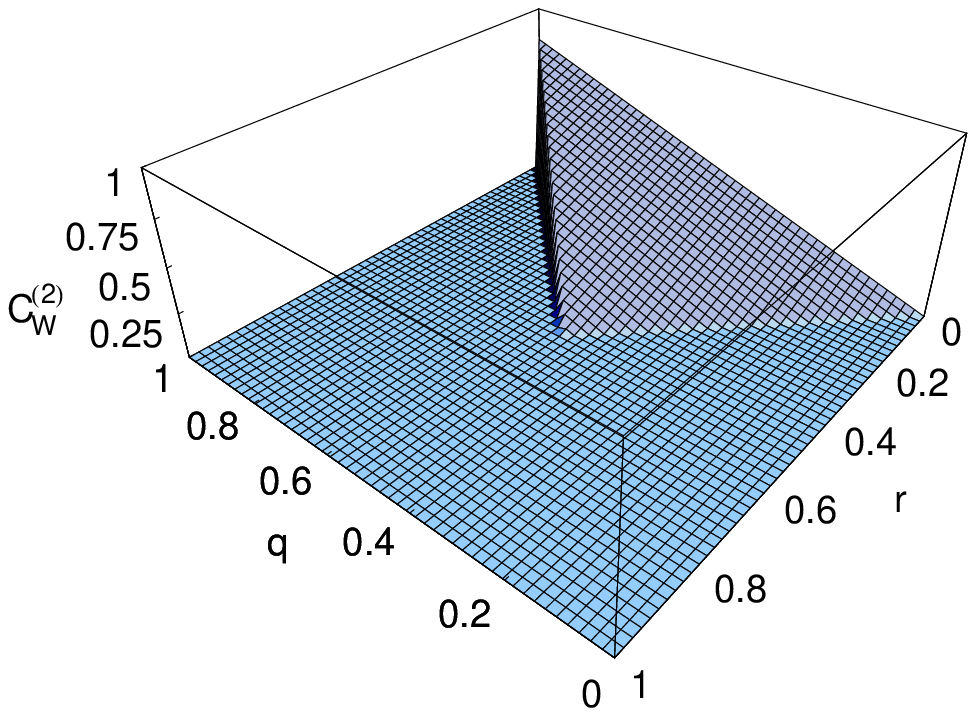}
\caption{Quantitative comparison of three bounds, the bound
(\ref{boundMB}) (plots (a)), our bound (\ref{boundTransp}) (plots
(b)), and the bound (\ref{boundBreuerW}) (plot (c)) for the class
of states (\ref{SO3}) for different values of the parameters $p$,
$q$, and $r$ ($C_{\mathrm{MB}}^{(i)}$, $C_{\mathrm{Tr}}^{(i)}$,
and $C_{W}^{(2)}$ stand for the functions appearing on the
right--hand sides of these bounds). Left plot is made for $p=0$,
while the right one for $q=0$ (in the case of (\ref{boundBreuerW})
only one bound is plotted for $q=0$ as the one for $p=0$ gives
always zero). Though the new bound (\ref{boundTransp}) provides
lower values it detects entanglement in places where the MB bound
as well as the bound (\ref{boundBreuerW}) give zero.}
\label{FigDetection2}
\end{figure}
\end{widetext}

In Fig. \ref{FigDetection2} one has quantitative comparison of
bounds (\ref{boundMB}) and (\ref{boundTransp}). As previously, for
completeness we also studied the bound (\ref{boundBreuerW}). It is
clear that even though our bound is worse in the sense that it
provides lower values, it works in the region in which the bound
(\ref{boundMB}) does detect anything.
%
\section{Measurable upper bounds on other concurrences}
\label{UpperBounds}

%

Here we discuss a possibility to generalize the upper bound
(\ref{boundDual}) on $C$ provided in Ref. \cite{Guo}. This bound
follows straightforwardly from the concavity of the function
$\sqrt{1-x^{2}}$. Here we point out that this reasoning may be
immediately extended into the whole class of concurrences
introduced firstly in Ref. \cite{Sinolecka} and then considered,
e.g., in Refs. \cite{Jap,Gour}. Further, we point out some
possibilities of extending this bounds to a more general class of
entanglement monotones. Moreover, we discuss these concurrences
and obtained upper bounds in the context of the so--called Schmidt
number of a density matrix \cite{Terhal}, which is also an
entanglement monotone.

Let us firstly introduce some notations. Let $\varrho$ be a given
$d\times d$ density matrix and $\lambda(\varrho)$ denote a
$d$--dimensional vector consisting of eigenvalues of $\varrho$
(hereafter denoted by $\lambda_{i}(\varrho)$). Then, let
$\sigma_{k}(x_{0},\ldots,x_{d-1})$ denote the so--called $k$th
elementary symmetric polynomial of $d$ arguments, that is
\begin{equation}\label{symmetric}
\sigma_{k}(x_{0},\ldots,x_{d-1})=\sum_{i_{1}<i_{2}<\ldots<i_{k}=0}^{d-1}x_{i_{1}}\ldots
x_{i_{k}}.
\end{equation}
Let us only mention that the first and last elementary symmetric
polynomials, i.e., the ones corresponding to $k=1$ and $k=d$ are
$x_{0}+\ldots +x_{d-1}$ and $x_{0}\cdot\ldots\cdot x_{d-1}$,
respectively. Now, following Refs. \cite{Sinolecka,Gour} (with
normalization adopted from Ref. \cite{Gour}) we can introduce the
class of $d$ concurrences of the form
\begin{equation}
C_{k}(\varrho)=\inf_{\{p_{i},\ket{\psi_{i}}\}}\sum_{i}p_{i}C_{k}(\ket{\psi_{i}})
\end{equation}
with $C_{k}$ defined for pure states in the following way
\begin{equation}\label{DefConck}
C_{k}(\ket{\psi})=h_{k}(\varrho_{r}) \qquad
(k=2,\ldots,d,\;r=A,B).
\end{equation}
Here, like before, $\varrho_{r}$ stands for one of the reductions
of $\varrho$ to a single--party state, while the functions $h_{k}$
are defined for $d\times d$ state $\rho$ as \cite{Gour}:
\begin{equation}\label{functionsh}
h_{k}(\rho)=\left(\frac{\sigma_{k}(\lambda(\rho))}{\sigma_{k}(\lambda(\mathbbm{1}_{d}/d))}\right)^{1/k}.
\end{equation}
First of all we need to emphasize that for purposes of the present
section the concurrence $C$ is denoted by $C_{2}$, however, with
different normalization. Namely, as it follows from Eqs.
(\ref{DefConck}) and (\ref{functionsh}), now
$C(\ket{\psi})=\sqrt{[d/(d-1)](1-\Tr\varrho_{r}^{2})}$ and
therefore it is normalized in such way that all the concurrences
$C_{k}$ $(k=3,\ldots,d)$ give one for maximally entangled state
$P_{+}^{(d)}$ irrespectively on $d$. For instance, the $d$th
concurrence, also called $G$--concurrence \cite{Gour} is given by
$C_{d}(\ket{\psi})\equiv
G(\ket{\psi})=d[\lambda_{0}(\rho_{r})\ldots\lambda_{d-1}(\rho_{r})]^{1/d}$.

Now we are prepared to proceed with our measurable upper bounds on
$C_{k}$. For this purpose let us notice that, as shown in Ref.
\cite{Gour}, the functions $h_{k}$ are concave, i.e., they satisfy
$h_{k}(p\rho_{1}+(1-p)\rho_{2})\geq
ph_{k}(\rho_{1})+(1-p)h_{k}(\rho_{2})$ for any density matrices
$\rho_{1}$ and $\rho_{2}$ and probability $0\leq p\leq 1$. Then,
denoting by $\{p_{i},\ket{\psi_{i}}\}$ an optimal ensemble
realizing $\varrho$ with respect to $C_{k}$, we are allowed to
write
\begin{eqnarray}\label{boundsGeneral}
C_{k}(\varrho)&=&\sum_{i}p_{i}C_{k}(\ket{\psi_{i}})\nonumber\\
&=&\sum_{i}p_{i}h_{k}(\varrho_{r}^{(i)})\nonumber\\
&\leq&h_{k}\left(\sum_{i}p_{i}\varrho_{r}^{(i)}\right)\nonumber\\
&=&h_{k}(\varrho_{r})\qquad (r=A,B).
\end{eqnarray}
To get the first equality we utilized Eq. (\ref{DefConck}), while
the inequality is a consequence of the aforementioned concavity of
$h_{k}$. Moreover, $\varrho_{r}^{(i)}$ and $\varrho_{r}$ $(r=A,B)$
denote reductions to the $r$th subsystems of pure states
$\ket{\psi_{i}}$ and $\varrho$, respectively.

Using explicit forms of $h_{f}$ (see (\ref{functionsh})), the
above bounds can be stated in the following form
\begin{eqnarray}\label{boundsOther}
&\displaystyle C_{2}(\varrho)\leq \sqrt{\frac{d}{d-1}\left(1-\Tr\varrho_{r}^{2}\right)}&\nonumber\\[2ex]
&\displaystyle C_{3}(\varrho)\leq \sqrt[3]{\frac{d^{2}}{(d-1)(d-2)}\left(1-3\Tr\varrho_{r}^{3}+2\Tr\varrho_{r}^{2}\right)}&\nonumber\\
&\vdots&\nonumber\\
&\displaystyle C_{d}(\varrho)\leq
d\left[\det(\varrho_{r})\right]^{1/d}.&
\end{eqnarray}
In the first inequality one recognizes the bound (\ref{boundDual})
provided in \cite{Guo}, however, with different normalization.

Let us notice that we can formalize the above considerations in a
little bit more general way. Let $\mathcal{C}$ be a function
defined for any pure state $\ket{\psi}\in \mathbbm{C}^{d}$ as
$\mathcal{C}(\ket{\psi})=h(\varrho_{A(B)})$ with $h$ denoting some
polynomial function of eigenvalues of $\varrho_{A(B)}$
($\varrho_{A(B)}$ denotes one of the subsystems of $\varrho$). On
knows \cite{NielsenMajor} that $\mathcal{C}$ can be an
entanglement monotone for pure states if and only if $h$ is a
Schur--concave function. That is, if it is a symmetric function
obeying the following condition
\begin{equation}
\lambda(\sigma_{1})\succ \lambda(\sigma_{2})\quad \Rightarrow
\quad h(\sigma_{1})\leq h(\sigma_{2})
\end{equation}
for any $\sigma_{1}$ and $\sigma_{2}$. The expression
$\lambda(\sigma_{1})\succ \lambda(\sigma_{2})$ means that
eigenvalues of $\sigma_{1}$ majorize those of $\sigma_{2}$, i.e.,
the inequalities $\sum_{i=1}^{k}\lambda_{i}(\sigma_{1})\geq
\sum_{i=1}^{k}\lambda_{i}(\sigma_{2})$ are satisfied for
$k=1,\ldots,d-1$ and equality holds for $k=d$ (note that due to
the fact that $\sigma_{1}$ and $\sigma_{2}$ are normalized this
condition is satisfied naturally).

Extension of $\mathcal{C}$ to all mixed states can be done using
the concept of convex roof. Now assuming that $h$ is a concave
function and following the reasoning given in Eq.
(\ref{boundsGeneral}), one immediately have that
$\mathcal{C}(\varrho)\leq h(\varrho_{A(B)})$. We know from Ref.
\cite{Brun} that any polynomial function of $\varrho_{A(B)}$ is
measurable as a mean value of some quantum observable on some
amount of copies of $\varrho_{A(B)}$ (the amount of copies depends
on the degree of a measured polynomial). Concluding, we have a
quite general statement saying that any such entanglement monotone
can be upper bounded by some measurable function of a given state.

Finally, it is interesting to discuss this bounds in the context
of another entanglement monotone, namely the Schmidt number
\cite{Terhal}. Let us recall that the Schmidt number of a
bipartite mixed state $\varrho$ is defined as
\begin{equation}\label{SchmidtNumber}
SN(\varrho)=\min_{\{p_{i},\ket{\psi_{i}}\}}\max_{i}
SR(\ket{\psi_{i}}),
\end{equation}
with $SR(\ket{\psi})$ being the so--called Schmidt rank of the
pure state $\ket{\psi}$, i.e., the number of nonzero Schmidt
coefficients in the Schmidt decomposition of $\ket{\psi}$. It is
clear that the concurrences $C_{k}$ $(k=1,\ldots,d)$ are directly
connected to the notion of Schmidt rank for pure states
\cite{Gour}. Namely, one sees from the definition of $C_{k}$ (Eqs.
(\ref{DefConck}) and (\ref{functionsh})) and the symmetric
polynomials $\sigma_{k}$ (\ref{symmetric}) that the Schmidt rank
of $\ket{\psi}$ is $l$ if and only if $C_{k}(\ket{\psi})\neq 0$
for $k\leq l$ and $C_{k}(\ket{\psi})=0$ for $k=l+1,\ldots,d$. This
reasoning can also be extended to the case of mixed states. More
precisely, we can prove that $SN(\varrho)= l$ iff
$C_{k}(\varrho)\neq 0$ for $k\leq l$ and $C_{k}(\varrho)=0$ for
$k=l+1,\ldots,d$. Notice that if for some $k$, $C_{k}(\varrho)=0$
then also $C_{l}(\varrho)=0$ for $l\geq k$. This is because if
$C_{k}(\varrho)=0$ then there exists such ensemble
$\mathcal{E}(\varrho)=\{p_{i},\ket{\psi_{i}}\}$ realizing
$\varrho$ that $SR(\ket{\psi_{i}})<k$ for each $i$. This, by
virtue of what was said above, means that
$C_{l}(\ket{\psi_{i}})=0$ for any $l\geq k$ and therefore the
ensemble $\mathcal{E}(\varrho)$ is also the optimal one for
$C_{l}$ with $l\geq k$. For a similar reason if
$C_{l}(\varrho)\neq 0$ then also $C_{k}(\varrho)\neq 0$ for $k\leq
l$. Since $C_{l}(\varrho)$ all the ensembles realizing $\varrho$
have to contain at least one pure state with Schmidt rank greater
or equal to $l$. Otherwise, $C_{l}$ would have to be zero. This
causes that all the concurrences $C_{k}$ with $k\leq l$ have to be
nonzero. Let us notice that the above two facts follow also from
the inequality given in \cite{Gour}, namely,
\begin{equation}
C_{2}^{2}(\ket{\psi})\geq C_{3}^{3}(\ket{\psi})\geq \ldots \geq
C_{d}^{d}(\ket{\psi}).
\end{equation}

By virtue of what was said it suffices to prove the statement that
$SN(\varrho)=l$ iff $C_{l}(\varrho)\neq 0$ and
$C_{l+1}(\varrho)=0$. For this purpose let us assume that
$SN(\varrho)=l$. Then, according to the definition of the Schmidt
number, one sees that there exists as ensemble
$\widetilde{\mathcal{E}}(\varrho)=\{q_{i},\ket{\varphi_{i}}\}$
realizing $\varrho$ for which $\max_{i}SR(\ket{\varphi_{i}})=l$.
This means on the one hand that $C_{k}(\ket{\varphi_{i}})=0$ for
any $i$ and $k=l+1,\ldots,d$. Consequently $C_{k}(\varrho)=0$ for
$k>l$ and the ensemble $\widetilde{\mathcal{E}}(\varrho)$ is the
optimal one for these concurrences. On the other hand, according
to the definition of the Schmidt number, all ensembles realizing
$\varrho$ must contain at least one pure state of which the
Schmidt rank is not less than $l$. This means that
$C_{k}(\varrho)\neq 0$ for $k\leq l$.

To deal with the opposite direction we assume that
$C_{l}(\varrho)\neq 0$ and $C_{l+1}(\varrho)\neq 0$. From the
latter we infer that there must exist an ensemble of $\varrho$ in
which all the pure states have the Schmidt rank at most $l$ and
therefore, according to (\ref{SchmidtNumber}), $SN(\varrho)\leq
l$. On the other hand, since $C_{l}(\varrho)\neq 0$ all the
ensembles must contain at least one state of which the Schmidt
rank is greater or equal to $l$. This means finally that
$SN(\varrho)\geq l$ and together with the previous fact that
$SN(\varrho)\leq l$ gives eventually $SN(\varrho)=l$.

The above discussion leads us to the conclusion that the bounds
(\ref{boundsGeneral}) can be also applied to bound the Schmidt
number of a given $\varrho$. Namely, if $h_{l}(\varrho_{r})=0$
(notice that then obviously $h_{k}(\varrho_{r})=0$ for
$k=l,\ldots,d$), then $C_{k}(\varrho)=0$ for any $k=l,\ldots,d$.
This, by virtue of the above statement, means that
$SN(\varrho)<l$. Therefore measuring upper bounds on the
concurrences $C_{k}$ we can also get information about the Schmidt
number of $\varrho$.

\section{Conclusion}
\label{Conclusions}
%

The general aim of the paper was to provide different ways of
improving existing measurable bounds on entanglement measures.
Measurable in the sense that the bounding functions can be written
as a mean value of some quantum observable on a single or many
copies of a state. Here we concentrated on those which can be
measured collectively on two copies of a state. Though being
rather harder to perform experimentally (one has to have several
identical uncorrelated copies of given $\varrho$ at the same time)
(see Ref. \cite{Enk2}) such bounds in general work better (in the
sense that they detect more states) than the ones based on linear
witnesses.

In the first step we have provided another proof of the recent
Mintert--Buchleitner bound \cite{MintertBuchleitner} on
concurrence and its multipartite generalization from Ref.
\cite{Aolita}. Then we have discussed possible improvements of
this bounds which could follow from the proof. Though we were not
able to provide a class of states for which the bound could be
improved (and still be measurable) it seems that it is still
interesting to investigate this approach.

Next, we have made an attempt to find other lower bounds on
concurrence that could also be measured on copies of a given
state. Firstly, relating the concurrence to the generalized
robustness of entanglement and utilizing properties of the latter
we have provided a way of bounding the former by mean values of
entanglement witnesses obeying $W\leq \mathbbm{1}_{d}$. This, to
some extent, can also be considered as a generalization of the
result of Ref. \cite{BreuerBound} (see Eq. \ref{boundBreuerW}).
Then, choosing appropriately observables satisfying the above
constraint we have obtained the whole class of bounds on
concurrence depending on an arbitrary positive map and measurable
on two copies of a state. Interestingly, using this approach one
can also provide a different proof of the MB bound. More
precisely, our method can be shown to reproduce the bound
(\ref{boundMB}) when one considers the reduction map. In
particular, we have investigated this bound for the transposition
map on the class of $4\otimes 4$ rotationally invariant states and
showed that, though rather less sharp, it works for regions for
which the Mintert--Buchleitner bound gives zero.

Finally, we have provided a general reasoning leading to upper
measurable bounds on the class of entanglement monotones. In
particular, we have discussed these bounds for the class of
concurrences provided in Refs. \cite{Sinolecka,Gour}.

Clearly, since as at least in the case of investigated states, the
presented method works better than the MB bound, it is worth to be
studied further. Namely, one has to try to optimize the utilized
entanglement witnesses (through the constants
$\alpha_{\rho}^{\Lambda}$) to improve the tightness of the bounds.
This in general seems to be a rather hard task. In the case of the
reduction map $R$ some steps towards optimization can be made
using the bound on fidelity \cite{Miszczak}, as it is pointed out
in the paper. The procedure then leads to the MB result.

On the other hand, one could investigate other positive maps than
the transposition one. For instance it would be particularly
interesting to study also the indecomposable maps as it could give
us measurable bounds detecting bound entangled states. This could
lead to experimental detection of bound entanglement through joint
measurements on two copies of a state. Finally, it would be
desirable to derive similar measurable bounds using the recent
entropic--like inequalities (\cite{RAJSPH,RAJS,RAJS2}), which were
shown to detect more entangled states than the entropic
inequalities (\ref{entropic}). Also, one could try to get similar
bounds for the other concurrences as for instance $C_{k}$.

Finally, let us stress that all of the presented bounds concern
mixed states, i.e., realistic situations in which the preparation
of entangled states is affected by noise, or where an initially
ideal entangled state undergoes dephasing. Experimental detection
of these bounds can be realized using the set up of Ref.
\cite{Schmid} for photons, but generalization to atom or ion pairs
are straightforward. Still, the presented bounds require
preparation of two (or in general more) identical copies of the
state of the system, and this process, is, of course, also subject
to experimental imperfections. Fortunately, as in the case of
Mintert--Buchleiner bound, all of our results are easily
generalized for pairs of different states. As demonstrated in Ref.
\cite{Schmid}, this allows to obtain conclusive information about
the entanglement of the state even in the worst case when the
second copy is maximally entangled (i.e., gives the main
contribution to the bound), provided the state of the system is
entangled sufficiently strongly.

\section{Acknowledgments}

We are grateful to Daniel Cavalcanti, Otfried G\"uhne, Julia
Stasi\'nska, Julio de Vicente, and Karol \.Zyczkowski for
discussions. This work was prepared under the financial support
from EU Programmes IP "SCALA" and STREP "NAMEQUAM", Spanish MEC
grants (FIS 2005-04627, FIS2008-00 and Consolider Ingenio 2010
"QOIT"), and Humboldt Foundation.

\end{document}